\pdfoutput=1
\documentclass[12pt]{article}

\usepackage{amsmath, amssymb, cite, color, footnote, graphicx, units, verbatim}
\usepackage[sort&compress,numbers,colon]{natbib}

\usepackage{tikz}
\usetikzlibrary{positioning,arrows,patterns}
\usetikzlibrary{decorations.pathmorphing}
\usetikzlibrary{decorations.markings}
\usetikzlibrary{calc}
\usetikzlibrary{shapes.misc}
\tikzset{
    photon/.style={decorate, decoration={snake}, draw=black, thick},
    fermionnoarrow/.style={draw=black, postaction={decorate}, thick},
    scalar/.style={draw=black, postaction={decorate}, thick, dashed},
    fermion/.style={draw=black, postaction={decorate},decoration={markings,mark=at position .55 with {\arrow{>}}}, thick},
    gluon/.style={decorate, draw=black, decoration={coil,amplitude=4pt, segment length=5pt}, thick},
    vertex/.style={draw,shape=circle,fill=black,minimum size=3pt,inner sep=0pt},
    cross/.style={cross out, thick, draw=black, fill=none, minimum size=2*(#1-\pgflinewidth), inner sep=0pt, outer sep=0pt}, cross/.default={2pt}
}

\newcommand{\lag}{\mathcal{L}}
\newcommand{\amp}{\mathcal{M}}
\newcommand{\lpint}{\mathcal{J}}

\newcommand{\kmix}{\epsilon}

\newcommand{\dm}{\delta m^2}

\newcommand{\set}[1]{\mathbb{#1}}
\newcommand{\secref}[1]{Section~\ref{#1}}
\newcommand{\appref}[1]{Appendix~\ref{#1}}
\newcommand{\modeqref}[1]{Eq.~\eqref{#1}}

\newcommand{\figref}[1]{Fig.~\ref{#1}}
\newcommand{\figsref}[1]{Figs.~\ref{#1}}
\newcommand{\abs}[1]{\lvert #1 \rvert}

\title{}

\begin{document}
\allowdisplaybreaks[1]

\begin{titlepage}

\begin{center}
{\LARGE\textbf{Novel Collider and Dark Matter\\ 
\vspace{.2cm}
Phenomenology of a Top-philic $Z'$}}
\\[10mm]
\vspace{.2cm}
{\normalsize\sc
Peter Cox$^{a,}$\footnote{\texttt{pcox@physics.unimelb.edu.au}},
Anibal D.~Medina$^{a,b,}$\footnote{\texttt{anibal.medina@cea.fr}},\\
Tirtha Sankar Ray$^{c,}$\footnote{\texttt{tirthasankar.ray@gmail.com}}
 and Andrew Spray$^{a,d,}$\footnote{\texttt{a.spray.work@gmail.com}}}
\\[5mm]
{\small\textit{$^{a}$ARC Centre of Excellence for Particle Physics at the Terascale,\\
School of Physics, The University of Melbourne, Victoria 3010, Australia\\
$^{b}$ Institut de Physiqu\'e Th\'eorique, Universit\'e Paris Saclay, CNRS, CEA, F-91191 Gif-sur-Yvette, France\\
$^{c}$Department of Physics and Center for Theoretical Studies,\\ 
		     Indian Institute of Technology,
		     Kharagpur 721302, India \\
$^{d}$ Center for Theoretical Physics of the Universe, \\Institute for Basic Science (IBS), Daejeon, 34051, Korea}
}
\end{center}

\vspace*{0.2cm}
\date{\today}

\begin{abstract}
  We consider extending the Standard Model by including an additional Abelian gauge group broken at low energies under which the right-handed top quark is the only effectively charged Standard Model fermion. The associated gauge boson $(Z')$ is then naturally top-philic and couples only to the rest of the SM particle content at loop-level or via kinetic mixing with the hypercharge gauge boson which is assumed to be small. Working at the effective theory level,  we demonstrate that such a minimal extension allows for an improved fitting of the $\sim 2\sigma$ excess observed in $t\bar{t}h$ searches at the LHC in a region of parameter space that satisfies existing collider constraints. We also present the reach of the LHC at 13 TeV in constraining the relevant region of parameter space. Additionally we show that within the same framework a suitably chosen fermion charged only under the exotic Abelian group can, in the region of parameter space preferred by the $\bar{t}th$ measurements, simultaneously explain the dark matter relic density and the $\gamma$-ray excess at the galactic center observed by the Fermi-LAT experiment. 

\end{abstract}

\end{titlepage}

\setcounter{footnote}{0}

\section{Introduction}

Augmenting the Standard Model (SM) at the TeV scale with an additional gauged abelian group $(U(1)')$ provides an economical handle to address various shortcomings of the SM. Various aspects of the phenomenology of the $U(1)'$ and the associated gauge boson, $Z'$, have been discussed in the literature; see for example~\cite{Langacker:2008yv} and references therein for a review. These unbroken Abelian groups naturally arise in the low scale phenomenology of Grand Unification Models with a GUT group of rank five or higher. They also arise from higher dimensional constructions where symmetry is broken by orbifolding.

In this paper we adopt a bottom up approach within this class of models without specifying any particular UV completion. We consider the possibility that the SM has an additional exotic abelian group with its own Higgs mechanism resulting in a massive $Z'$ gauge boson. We will assume that only the right-handed top quark is charged under this exotic $U(1)'$ making the $Z'$ top-philic\footnote{For models where the $Z'$ preferentially couples to the bottom quark see \cite{Grossmann:2010wm}; while for models where the $Z'$ couples to all right-handed quarks, see \cite{Ko:2012hd,Ko:2012ud}.}. The $Z'$ can have kinetic mixing with the hypercharge gauge boson of the SM providing an additional portal for states in the SM and exotic sectors to communicate with each other. Such a top-philic $Z'$ has been discussed in the literature in the context of dark matter (DM) \cite{D'Hondt:2015jbs} and galactic $\gamma$-ray lines \cite{Jackson:2009kg,Jackson:2013rqp} as well as in the context of vacuum stability \cite{DiChiara:2014wha}, and can be motivated from models of composite/RS dark matter~\cite{Jackson:2009kg}. Such a model also requires additional chiral fermions for anomaly cancellation. We will assume that additional fermions in the exotic sector take care of this but are heavy enough to be decoupled from our effective theory at low energies. Our goal in this paper is to study the possibility that a minimal extension of the SM which includes a top-philic $Z'$ is able to address some recent signals that are hard to explain within the SM. We will try to reconcile some observed experimental tension in the LHC data with the apparently uncorrelated issue of dark matter and the galactic $\gamma$-ray excess using such a construction.

The ATLAS and CMS collaborations have recently released their combined Higgs measurements using the complete Run I dataset from the LHC~\cite{ATLAS-CONF-2015-044,CMS:2015kwa}. While overall the results show good agreement with Standard Model expectations, they observe a slight excess in the case of Higgs production in association with a pair of top quarks. This process is of particular importance since it allows, in principle, for a direct measurement of the top quark Yukawa coupling, which otherwise only enters via loops in the Higgs production and decays into massless gauge bosons. Though the SM cross-section for this process is small ($\sigma_{SM}\sim 130\,$fb at $\sqrt{s}=8\,$TeV) and thus measuring its rate challenging at the LHC Run I, interestingly the combined measurement by ATLAS and CMS gives a $t\bar{t}h$ signal strength relative to the SM cross-section of $\mu=\sigma/\sigma_{SM}=2.3^{+0.7}_{-0.6}$ for a 125~GeV Higgs boson. This deviation from the SM value ($\mu=1$) is driven by the multi-lepton channels and in particular the CMS same-sign dilepton channel, which gives a best fit value of $\mu=5.3^{+2.1}_{-1.8}$. Overall, compared to the SM expectation, the observed excess is equivalent to a 2.3-standard-deviation ($\sigma$) upward fluctuation. We will show that an additional contribution arising from the production of the $Z'$ in association with top quarks ($t\bar{t}Z'$) could provide an explanation for the observed excess.  (For previous theoretical studies of this excess, see~\cite{Bhattacherjee:2015sia,vonBuddenbrock:2015ema,Feldmann:2015zwa,Huang:2015fba,Chen:2015jmn}.) We stress that a more precise measurement of this Higgs production mechanism will be one of the key objectives at the LHC Run II, making our study relevant for the near future.

On the other hand an intriguing excess of $\gamma$-rays from the galactic center has been observed in the Fermi-LAT data~\cite{Daylan:2014rsa, Calore:2014xka}. The observed excess is contingent on the possible large uncertainties in the astrophysical $\gamma$-ray background \cite{Calore:2014nla} and may in the end have an astrophysical origin. However, it can be well fitted by the continuum photon spectrum provided by DM annihilating in the center of the Milky Way, within the usual thermal weakly interacting massive particle paradigm and for the standard DM relic density profiles. The excess is observed in the energy range of 10 MeV to 300 GeV with a peak around 3 GeV. In the case of our $Z'$ extension to the SM, the Higgs mechanism in the exotic sector that breaks the $U(1)'$ may leave an unbroken $Z_2$ under which all the SM states are even. Thus the lightest $Z_2$ odd state in this exotic sector can become a viable dark matter candidate. The DM can then annihilate to SM states via the $Z'$, potentially leading to an observable $\gamma$-ray signal.

To summarize, we consider a top-philic $Z'$ in the mass range 150-450 GeV and determine the phenomenologically relevant production mechanisms and decay widths. We translate the constraints from electroweak precision tests and direct searches at colliders and identify the region of parameter space that results in an improved fit of the $t\bar{t}h$ measurements. Furthermore we show that if the exotic sector contains a stabilized Dirac fermion, it can provide a simultaneous solution to the DM relic density constraints and the galactic center excess that is also compatible with the fit to the $t\bar{t}h$ data. We also present the reach of the LHC run at $\sqrt{s}=13\,$TeV to explore the relevant regions of parameter space.

This paper is organized as follows: in Sections~\ref{sec:model} and~\ref{sec:prod_decay} we introduce the effective theory for the $Z'$ and briefly discuss the main phenomenological aspects. Then in Section~\ref{sec:collider_searches} we discuss the improved fit to the measured $t\bar{t}h$ signal strength and present the current and projected constraints from collider searches. Finally in Section~\ref{sec:dmge} we discuss the dark matter and galactic center $\gamma$-ray excess within this framework before concluding in Section~\ref{sec:conc}.

\section{Model} \label{sec:model}

In this section we introduce the effective Lagrangian obtained by integrating out the exotic Higgs sector responsible for breaking the $U(1)'$ group and giving mass to the associated gauge boson $Z'_\mu$. The low-energy effective theory for the top-philic $Z'$ and a Dirac fermion $\chi$, neutral under all SM gauge symmetries but charged under $U(1)'$, is given by~\cite{Jackson:2009kg}
\begin{equation}
  \begin{split}
    \lag & = \lag_{SM} - \frac{1}{4} Z'_{\mu\nu} Z'{}^{\mu\nu} - \frac{1}{2} \, \kmix \, Z'_{\mu\nu} B^{\mu\nu} + \frac{1}{2} \, m_{Z'}^2 \, Z'_\mu Z'{}^\mu \\
    & \quad + g_t \, Z'_\mu \, \bar{t} \gamma^\mu P_R t + \bar{\chi} \gamma^\mu \bigl( i \partial_\mu + g_\chi Z'_\mu \bigr) \chi - m_\chi \, \bar{\chi}\chi \,.
  \end{split}
\label{eq:lagrangian}\end{equation}
$P_R$ is the usual projection operator, which ensures that the $Z'$ couples only to the right-handed top.  $B_{\mu\nu}$ is the hyper-charge field strength, and with some abuse of notation we have used $Z'$ to represent both the gauge and mass eigenstates for the new vector.  In the presence of non-zero kinetic mixing $\kmix$, the two bases are related as~\cite{0608068,1011.3300}
\begin{equation}
  (Z'_\mu)^{mass} = (Z'_\mu)^{gauge} + \mathcal{O} (\kmix) \,, \quad \text{and} \quad (m_{Z'}^2)^{pole} = m_{Z'}^2 + \mathcal{O} (\kmix \, m_Z^2) \,. 
\end{equation}
Precision electroweak constraints bound $\kmix \lesssim10^{-1}$--$10^{-2}$ for $m_{Z'} \in [150, 450]$~GeV~\cite{1006.0973} (with the bound weakening as $m_{Z'}$ increases) so these corrections are small. 

The $U(1)'$ as given in \modeqref{eq:lagrangian} is clearly anomalous, since the right-handed top is the only SM state with non-zero $U(1)'$ charge.  We assume that all the anomalies are cancelled by spectator fermions at a scale $\Lambda_{UV}$.  The details of the $U(1)'$ breaking can also be deferred to that scale. Cancellation of mixed anomalies and a viable top Yukawa coupling imply the existence of additional colored resonances at the cutoff scale with  present mass limits at $\gtrsim 900$ GeV \cite{Chatrchyan:2013uxa, Aad:2014efa}.  We expect $\Lambda_{UV} \gtrsim 900 \ \text{GeV}\gg m_{Z'}, m_\chi$.  This justifies studying the phenomenology within the effective theory. 

The mixing parameter $\kmix$ depends on the details of the UV completion, so we treat it as a free parameter of the low energy theory.  However, loops of top quarks in the effective theory generate a kinetic mixing of size
\begin{equation}
  \kmix^{top-loops} \approx \frac{2}{3} \, \frac{N_c g_t g_1}{(4\pi)^2} \, \log \frac{\Lambda_{UV}^2}{m_t^2} \,,
\end{equation}
where $\Lambda_{UV}$ is the cut-off of the theory.  For $\Lambda_{UV} \approx 1$~TeV, this contribution is $\kmix \approx g_t \times 10^{-2}$, safely below the precision constraints quoted above for $g_t < 1$.  UV physics can increase or (with some tuning) decrease the magnitude of $\kmix$.

Similarly, top quark loops and UV physics can generate the operator $Z'_\mu \, (H^\dagger \mathcal{D}^\mu H - (\mathcal{D}^\mu H)^\dagger H)$, which in the low energy theory corresponds to a mass mixing 
\begin{equation}
  \lag \supset \dm \, Z'_\mu Z^\mu \,.
\label{eq:defdm}\end{equation}
Top quark loops alone result in
\begin{equation}
  (\delta m^2)^{top-loop} \approx \frac{1}{2} \, m_t^2 \, \frac{N_c g_t g_2}{(4\pi)^2} \, \log \frac{\Lambda_{UV}^2}{m_t^2} \approx g_t (30 \text{ GeV})^2 \,,
\end{equation}
where we have again taken $\Lambda_{UV} \approx 1$~TeV.  This can have important effects on the $Z'$ coupling to nucleons~\cite{1402.1173,1411.3342}, and also leads to corrections to the $Z$ mass.  The latter leads to the bound~\cite{1006.0973}
\begin{equation}
  \abs{\delta m^2} \lesssim 0.007 \, m_Z \, m_{Z'} \,,
\label{eq:dm2limit}\end{equation}
or $m_{Z'}/g_t \gtrsim 1.2$~TeV.  As was the case for the kinetic mixing, this term depends on the UV completion and so is a free parameter in the low energy theory.  In particular, an $\mathcal{O}(10\%)$ tuning with the UV completion would suppress the mixing sufficiently to allow $m_{Z'}/g_t \sim 100$~GeV.

Our theory respects an accidental $\set{Z}_2$ symmetry under which $\chi$ is odd while all other fields are even. This global discrete symmetry ensures that $\chi$ is stable on cosmological scales and a viable dark matter candidate.  The $\set{Z}_2$ is naturally interpreted as the residue of breaking the $U(1)'$ gauge group, which ensures that it is not broken by higher-dimensional operators~\cite{Rothstein:1992rh,1506.05107,1506.06767}.  The $Z'$ serves as the mediator between the visible and dark sectors, so that the DM is also top-philic.

\section{The properties of the $Z'$} \label{sec:prod_decay}

In this section we briefly outline the main properties of the $Z'$ gauge boson, in particular the important production and decay mechanisms.

\subsection{$Z'$ Production}

The production of a top-philic vector resonance at the LHC was previously considered in detail in Ref.~\cite{1410.6099} for resonance masses above the $t\bar{t}$ threshold. Here we briefly summarize the production mechanisms most relevant for LHC searches while focusing in more detail on the lower mass region. 

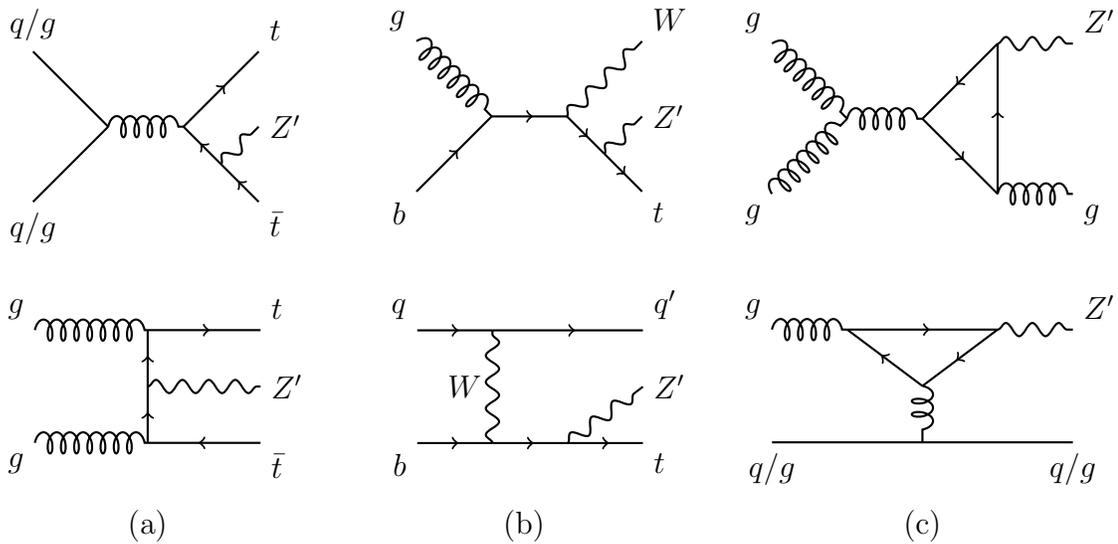
\begin{figure}
  \centering
  \parbox{0.3\textwidth}
  {
    \centering
    \begin{tikzpicture}[node distance=0.5cm and 0.5cm]
      \coordinate (v1);
      \coordinate[above left = of v1] (va);
      \coordinate[below left = of v1] (vb);
      \coordinate[above left = of va, label=above:$q/g$] (i1);
      \coordinate[below left = of vb, label=below:$q/g$] (i2);
      \coordinate[right = of v1] (vc);
      \coordinate[right = of vc] (v2);
      \coordinate[above right = of v2] (v3);
      \coordinate[above right = of v3, label=above right:{$t$}] (o1);
      \coordinate[below right = of v2] (v4);
      \coordinate[above right = of v4, label=right:{$Z'$}] (o2);
      \coordinate[below right = of v4, label=below right:{$\bar{t}$}] (o3);
      \draw[fermionnoarrow] (i1) -- (v1);
      \draw[fermionnoarrow] (v1) -- (i2);
      \draw[gluon] (v1) -- (v2);
      \draw[fermion] (v2) -- (o1);
      \draw[fermion] (o3) -- (v4);
      \draw[fermion] (v4) -- (v2);
      \draw[photon] (v4) -- (o2);
    \end{tikzpicture}\vspace{12pt}
    \begin{tikzpicture}[node distance=0.75cm and 1.5cm]
      \coordinate (v1);
      \coordinate[above = of v1] (v2);
      \coordinate[below = of v1] (v3);
      \coordinate[left = of v2, label=above left:{$g$}] (i1);
      \coordinate[left = of v3, label=below left:{$g$}] (i2);
      \coordinate[right = of v2, label=above right:{$t$}] (o1);
      \coordinate[right = of v1, label=right:{$Z'$}] (o2);
      \coordinate[right = of v3, label=below right:{$\bar{t}$}] (o3);
      \coordinate[below = of v3, label = below:{(a)}] (v4);
      \draw[gluon] (i1) -- (v2);
      \draw[gluon] (i2) -- (v3);
      \draw[fermion] (o3) -- (v3);
      \draw[fermion] (v3) -- (v1);
      \draw[fermion] (v1) -- (v2);
      \draw[fermion] (v2) -- (o1);
      \draw[photon] (v1) -- (o2);
    \end{tikzpicture}
  }
  \parbox{0.3\textwidth}
  {
    \centering
    \begin{tikzpicture}[node distance=0.5cm and 0.5cm]
      \coordinate (v1);
      \coordinate[above left = of v1] (va);
      \coordinate[below left = of v1] (vb);
      \coordinate[above left = of va, label=above left:$g$] (i1);
      \coordinate[below left = of vb, label=below left:$b$] (i2);
      \coordinate[right = of v1] (vc);
      \coordinate[right = of vc] (v2);
      \coordinate[above right = of v2] (v3);
      \coordinate[above right = of v3, label=above right:{$W$}] (o1);
      \coordinate[below right = of v2] (v4);
      \coordinate[above right = of v4, label=right:{$Z'$}] (o2);
      \coordinate[below right = of v4, label=below right:{$t$}] (o3);
      \draw[gluon] (i1) -- (v1);
      \draw[fermion] (i2) -- (v1);
      \draw[fermion] (v1) -- (v2);
      \draw[fermion] (v2) -- (v4);
      \draw[fermion] (v4) -- (o3);
      \draw[photon] (v2) -- (o1);
      \draw[photon] (v4) -- (o2);
    \end{tikzpicture}\vspace{16pt}
    \begin{tikzpicture}[node distance=0.75cm and 1cm]
      \coordinate[label=left:{$W$}] (v1);
      \coordinate[above = of v1] (v2);
      \coordinate[below = of v1] (v3);
      \coordinate[left = of v2, label=above left:{$q$}] (i1);
      \coordinate[left = of v3, label=below left:{$b$}] (i2);
      \coordinate[right = of v2] (va);
      \coordinate[right = of va, label=above right:{$q'$}] (o1);
      \coordinate[right = of v3] (vb);
      \coordinate[right = of vb, label=below right:{$t$}] (o3);
      \coordinate[below = of v3, label=below right:{(b)}] (v4);
      \coordinate[above = of o3, label=right:{$Z'$}] (o2);
      \draw[fermion] (i1) -- (v2);
      \draw[fermion] (i2) -- (v3);
      \draw[fermion] (v3) -- (vb);
      \draw[fermion] (vb) -- (o3);
      \draw[photon] (v3) -- (v2);
      \draw[fermion] (v2) -- (o1);
      \draw[photon] (vb) -- (o2);
    \end{tikzpicture}
  }
  \parbox{0.3\textwidth}
  {
    \centering
    \begin{tikzpicture}[node distance=1cm and 1cm]
      \coordinate (v1);
      \coordinate[above left = of v1, label=above left:$g$] (i1);
      \coordinate[below left = of v1, label=below left:$g$] (i2);
      \coordinate[right = of v1] (v2);
      \coordinate[above right = of v2] (v3);
      \coordinate[right = of v3, label=above right:{$Z'$}] (o1);
      \coordinate[below right = of v2] (v4);
      \coordinate[right = of v4, label=below right:{$g$}] (o3);
      \draw[gluon] (i1) -- (v1);
      \draw[gluon] (i2) -- (v1);
      \draw[gluon] (v1) -- (v2);
      \draw[fermion] (v2) -- (v4);
      \draw[fermion] (v4) -- (v3);
      \draw[fermion] (v3) -- (v2);
      \draw[photon] (v3) -- (o1);
      \draw[gluon] (v4) -- (o3);
    \end{tikzpicture}\vspace{18pt}
    \begin{tikzpicture}[node distance=0.75cm and 1cm]
      \coordinate (v1);
      \coordinate[below = of v1] (v2);
      \coordinate[above left = of v1] (v3);
      \coordinate[above right = of v1] (v4);
      \coordinate[left = of v3, label=above left:{$g$}] (i1);
      \coordinate[left = of v2] (va);
      \coordinate[left = of va, label=below:{$q/g$}] (i2);
      \coordinate[right = of v4, label=above right:{$Z'$}] (o1);
      \coordinate[right = of v2] (vb);
      \coordinate[right = of vb, label=below:{$q/g$}] (o2);
      \coordinate[below = of v2, label=below:{(c)}] (vc);
      \draw[fermionnoarrow] (i2) -- (o2);
      \draw[gluon] (i1) -- (v3);
      \draw[gluon] (v1) -- (v2);
      \draw[fermion] (v1) -- (v3);
      \draw[fermion] (v3) -- (v4);
      \draw[fermion] (v4) -- (v1);
      \draw[photon] (v4) -- (o1);
    \end{tikzpicture}
  }
  \caption{$Z'$ production channels at the LHC.  (a): Leading contributions to $ttZ'$ associated production. (b): Examples of associated production with $b$ quark initial states. (c): Top-loop production of $Z'j$.}\label{fig:prod}
\end{figure}

A top-philic $Z'$ has several production mechanisms which are potentially of interest at the LHC. Firstly there are the tree-level production modes which involve the production of the $Z'$ in association with at least one additional top quark. Of these, production in association with a pair of top quarks ($pp\,{\rightarrow}\,Z'+t\bar{t}$) has the largest cross-section and, as we shall show, is of particular interest for searches at the LHC. There are also additional modes with reduced cross-sections such as $pp\,{\rightarrow}\,Z'+tj$ and $pp\,{\rightarrow}\,Z'+tW$ which require the presence of a $b$-quark in the initial state.  These processes are shown in \figref{fig:prod}\,(a) and~(b) respectively.

The $Z'$ can also be produced via the loop-induced process $pp\,{\rightarrow}\,Z'+\,$jets. In the case of small $Z-Z'$ mixing, the $Z'$ coupling to the light quarks is highly suppressed and tree-level contributions to this process can be neglected. Nevertheless, this process still leads to a larger cross-section than the tree-level associated production mechanisms considered above. The absence of a tree-level contribution does however lead to some subtleties even when computing the leading-order contribution to the cross-section. To begin with the Landau-Yang theorem~\cite{Yang:1950rg,Landau:1948kw} ensures that exclusive ($Z'+0\,$jet) production vanishes when the $Z'$ is produced on-shell. The leading contribution to the cross-section therefore comes from the $Z'+1\,$jet process where there is an additional parton in the final state, which is shown in \figref{fig:prod}\,(c). However, in many cases we shall be interested in experimental analyses which are inclusive in the number of jets, meaning we are forced to consider the contribution of the phase space region where the additional jet is soft and would normally be unresolved. Of course we cannot compute the cross-section in this region of phase space without properly taking into account resummation effects. While a detailed treatment of this process would certainly be interesting it is beyond the scope of this paper. Instead we shall follow the approach taken in~\cite{1410.6099} and impose the following loose cuts on the additional jet: $p_T>20\,$GeV, $|\eta|<6.0$.

\begin{figure}
  \centering
  \includegraphics[width=0.7\textwidth]{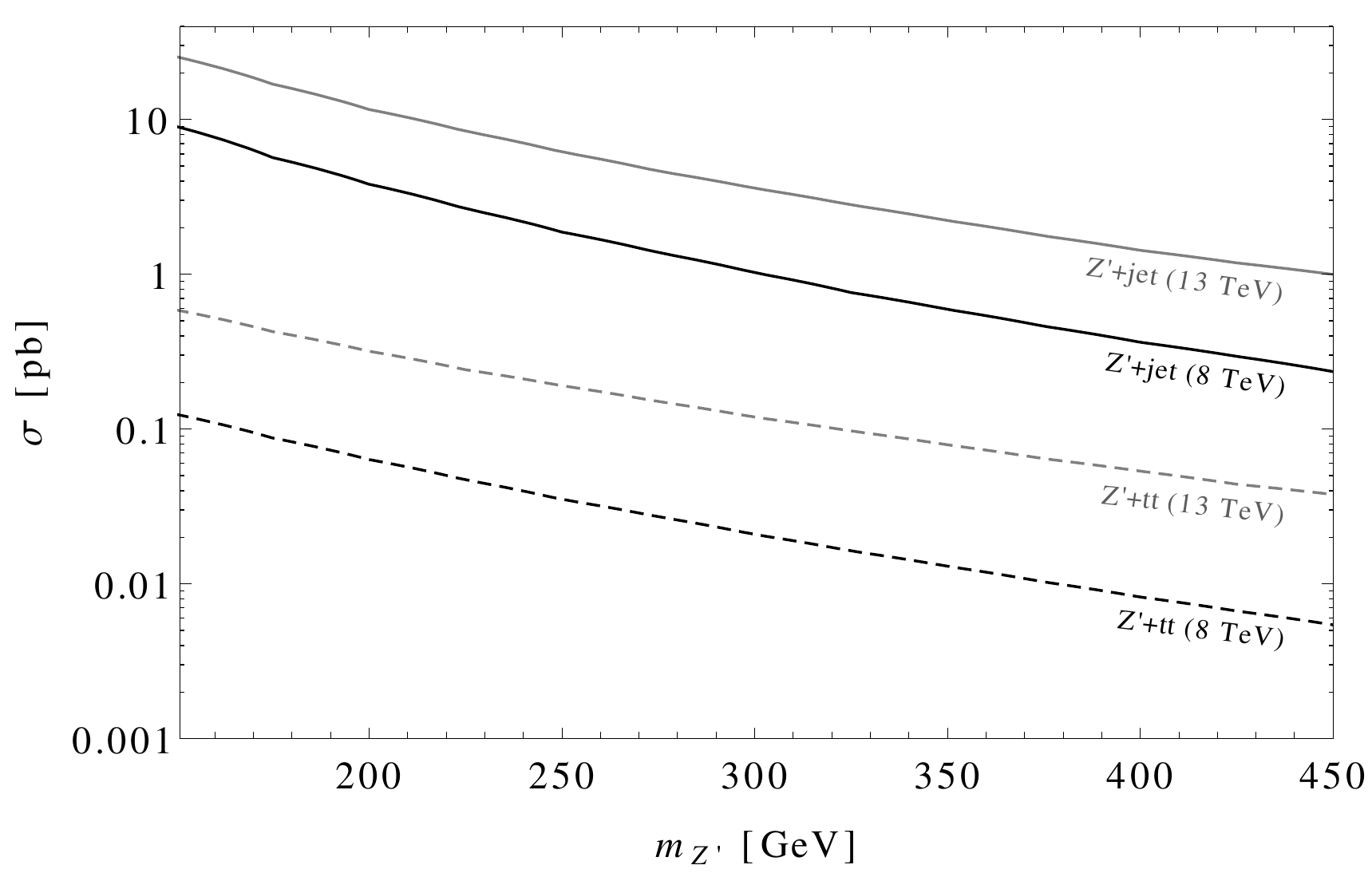}
  \caption{$Z'$ production cross-sections as a function of mass. The dashed lines correspond to $Z'$ production in association with top quarks, while the solid lines correspond to $Z'$+jets. The black (grey) lines are for $\sqrt{s}=8$ (13) TeV. We have set $g_t=0.5$ and $\epsilon=0$.} \label{fig:Z'xsec}
\end{figure}

The cross-sections for the $ttZ'$ and $Z'+\,$jets production modes are then shown in \figref{fig:Z'xsec} as a function of the $Z'$ mass. The cross-sections have been computed for $\sqrt{s}=8,\,13$~TeV at leading order (one-loop) in the case of $Z'+\,$jets and at NLO for the $ttZ'$ associated production using {\tt MadGraph5\_aMC@NLO}~\cite{1405.0301} and the {\tt NNPDF2.3} parton distribution functions~\cite{1207.1303}.

\subsection{$Z'$ Decays}

The phenomenology in this model, both at colliders and for the dark matter, is strongly dependent on the decays of the $Z'$. We are most interested in the case $m_{Z'} < 2m_\chi$, so that the $Z'$ has only visible collider decays.  The case $m_{Z'} > 2m_t$ was previously studied in~\cite{1410.6099}, so we will largely focus on the complementary scenario. The $Z'$ then has no tree-level two-body decays that are not suppressed by small mixing with the $Z$. The result is that multiple decay modes can be relevant, with a much richer phenomenology. 

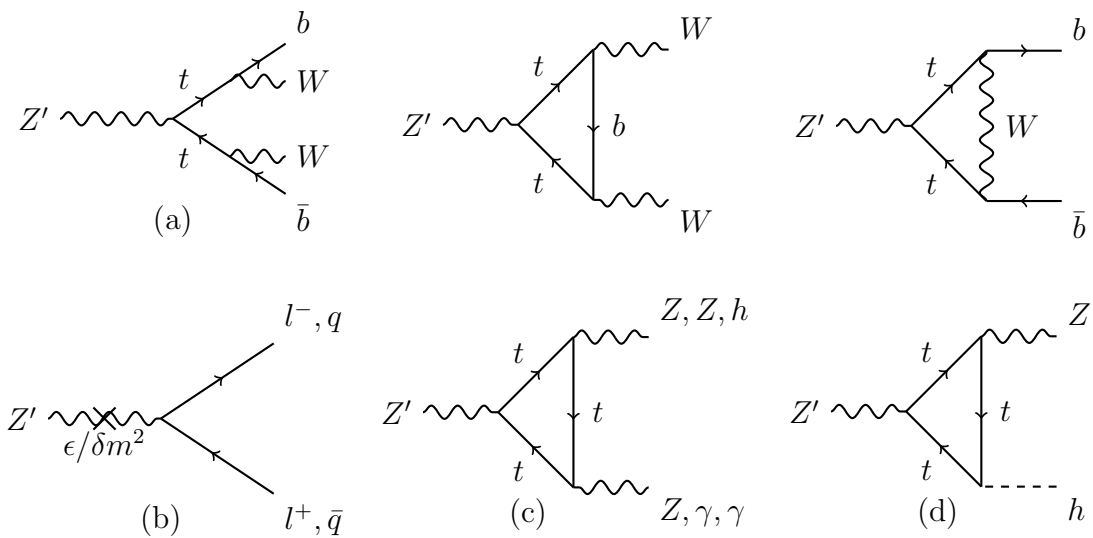
\begin{figure}
  \centering
  \parbox{0.3\textwidth}
  {
    \centering
    \begin{tikzpicture}[node distance=0.5cm and 0.75cm]
      \coordinate (v1);
      \coordinate[left = of v1] (va);
      \coordinate[left = of va, label=left:{$Z'$}] (i1);
      \coordinate[above right = of v1] (v2);
      \coordinate[below right = of v1] (v3);
      \coordinate[above right = of v2, label=above right:{$b$}] (o1);
      \coordinate[right = of v2, label=right:{$W$}] (o2);
      \coordinate[right = of v3, label=right:{$W$}] (o3);
      \coordinate[below right = of v3, label=below right:{$\bar{b}$}] (o4);
      \coordinate[below = of v1] (vb);
      \coordinate[below = of vb, label=below:{(a)}] (vc);
      \draw[photon] (i1) -- (v1);
      \draw[fermion] (o4) -- (v3);
      \draw[fermion] (v3) -- (v1) node[midway, below left=0cm] {$t$};
      \draw[fermion] (v1) -- (v2) node[midway, above left=0cm] {$t$};
      \draw[fermion] (v2) -- (o1);
      \draw[photon] (v3) -- (o3);
      \draw[photon] (v2) -- (o2);
    \end{tikzpicture}\vspace{16pt}
    \begin{tikzpicture}[node distance=0.5cm and 0.75cm]
      \coordinate (v1);
      \coordinate[left = of v1, label=below:{$\kmix/\delta m^2$}] (va);
      \coordinate[left = of va, label=left:{$Z'$}] (i1);
      \coordinate[above right = of v1] (vb);
      \coordinate[above right = of vb, label=above right:{$l^-,q$}] (o1);
      \coordinate[below right = of v1] (vc);
      \coordinate[below right = of vc, label=below right:{$l^+,\bar{q}$}] (o2);
      \coordinate[below = of v1] (vd);
      \coordinate[below = of vd, label=below:{(b)}] (ve);
      \draw[photon] (i1) -- (v1);
      \draw[fermion] (o2) -- (v1);
      \draw[fermion] (v1) -- (o1);
      \draw (va) node[cross=5pt] {};
    \end{tikzpicture}
  }
  \parbox{0.3\textwidth}
  {
    \centering
    \begin{tikzpicture}[node distance=1cm and 1cm]
      \coordinate (v1);
      \coordinate[left = of v1, label=left:{$Z'$}] (i1);
      \coordinate[above right = of v1] (v2);
      \coordinate[below right = of v1] (v3);
      \coordinate[right = of v2, label=above right:{$W$}] (o1);
      \coordinate[right = of v3, label=below right:{$W$}] (o4);
      \draw[photon] (i1) -- (v1);
      \draw[photon] (o4) -- (v3);
      \draw[fermion] (v3) -- (v1) node[midway, below left=0cm] {$t$};
      \draw[fermion] (v1) -- (v2) node[midway, above left=0cm] {$t$};
      \draw[photon] (v2) -- (o1);
      \draw[fermion] (v2) -- (v3) node[midway, right=0.1cm] {$b$};
    \end{tikzpicture}\vspace{16pt}
    \begin{tikzpicture}[node distance=1cm and 1cm]
      \coordinate (v1);
      \coordinate[left = of v1, label=left:{$Z'$}] (i1);
      \coordinate[above right = of v1] (v2);
      \coordinate[below right = of v1] (v3);
      \coordinate[right = of v2, label=above right:{$Z,Z,h$}] (o1);
      \coordinate[right = of v3, label=below right:{$Z,\gamma,\gamma$}] (o4);
      \coordinate[left = of v3, label=below right:{(c)}] (va);
      \draw[photon] (i1) -- (v1);
      \draw[photon] (o4) -- (v3);
      \draw[fermion] (v3) -- (v1) node[midway, below left=0cm] {$t$};
      \draw[fermion] (v1) -- (v2) node[midway, above left=0cm] {$t$};
      \draw[photon] (v2) -- (o1);
      \draw[fermion] (v2) -- (v3) node[midway, right=0.1cm] {$t$};
    \end{tikzpicture}
  }
  \parbox{0.3\textwidth}
  {
    \centering
    \begin{tikzpicture}[node distance=1cm and 1cm]
      \coordinate (v1);
      \coordinate[left = of v1, label=left:{$Z'$}] (i1);
      \coordinate[above right = of v1] (v2);
      \coordinate[below right = of v1] (v3);
      \coordinate[right = of v2, label=above right:{$b$}] (o1);
      \coordinate[right = of v3, label=below right:{$\bar{b}$}] (o4);
      \draw[photon] (i1) -- (v1);
      \draw[fermion] (o4) -- (v3);
      \draw[fermion] (v3) -- (v1) node[midway, below left=0cm] {$t$};
      \draw[fermion] (v1) -- (v2) node[midway, above left=0cm] {$t$};
      \draw[fermion] (v2) -- (o1);
      \draw[photon] (v3) -- (v2) node[midway, right=0.1cm] {$W$};
    \end{tikzpicture}\vspace{16pt}
    \begin{tikzpicture}[node distance=1cm and 1cm]
      \coordinate (v1);
      \coordinate[left = of v1, label=left:{$Z'$}] (i1);
      \coordinate[above right = of v1] (v2);
      \coordinate[below right = of v1] (v3);
      \coordinate[right = of v2, label=above right:{$Z$}] (o1);
      \coordinate[right = of v3, label=below right:{$h$}] (o4);
      \coordinate[left = of v3, label=below right:{(d)}] (va);
      \draw[photon] (i1) -- (v1);
      \draw[scalar] (o4) -- (v3);
      \draw[fermion] (v3) -- (v1) node[midway, below left=0cm] {$t$};
      \draw[fermion] (v1) -- (v2) node[midway, above left=0cm] {$t$};
      \draw[photon] (v2) -- (o1);
      \draw[fermion] (v2) -- (v3) node[midway, right=0.1cm] {$t$};
    \end{tikzpicture}
  }
  \caption{Decay modes of the $Z'$. (a): Tree-level decay to $t^{(\ast)}\bar{t}^{(\ast)}$.  (b): Tree-level decay from mass and kinetic mixing to leptons and light quarks.  (c): UV-finite loop decays.  (d): UV-divergent loop decays.}\label{fig:decay}
\end{figure}

We can classify the decays at zero mixing into three groups. First are three- and four-body tree-level decays, of which $Z' \to tW^-\bar{b}/\bar{t}W^+b$ is most important; see \figref{fig:decay}\,(a). Second are UV-finite loop-level decays, with final states $VV$ (for $V$ a SM gauge boson) and $h\gamma$; these are shown in \figref{fig:decay}\,(c).  Finally, there are UV-sensitive loop-level decays shown in \figref{fig:decay}\,(d): $Z' \to Zh$ and $Z'\to b\bar{b}$.\footnote{Strictly, there are also decays to all down-type quarks suppressed by off-diagonal elements of the CKM matrix.} These decays involve divergent loops, but in the full theory the divergences must cancel as there is no corresponding tree-level counter-term. They then depend on both the scale and the details of the UV completion.  Since these loops are closely related to the anomaly of \modeqref{eq:lagrangian}, we truncate the divergent integrals at the UV cut-off $\Lambda_{UV} \sim 900$~GeV. This means that the effective $Z'b\bar{b}$ and $Z'Zh$ couplings are enhanced by $\sim \log \Lambda_{UV}^2/m_{Z'}^2 \sim $~2--3, and generically dominate over the other loop-mediated decays.

In the presence of non-zero mass and kinetic mixing, there are additional decays to leptons and light quarks, shown in \figref{fig:decay}\,(b).  There will also be interference between the loop and mixing contributions for the decays to $WW$, $b\bar{b}$ and $Zh$. In the latter two, the loop contributions always dominate due to the logarithmic enhancement noted above. For the $WW$ decay, the interference is constructive for $\kmix > 0$ as defined in \modeqref{eq:lagrangian} and for $\dm > 0$ as defined in \modeqref{eq:defdm}.

\begin{figure}
  \centering
  \includegraphics[width=0.6\textwidth]{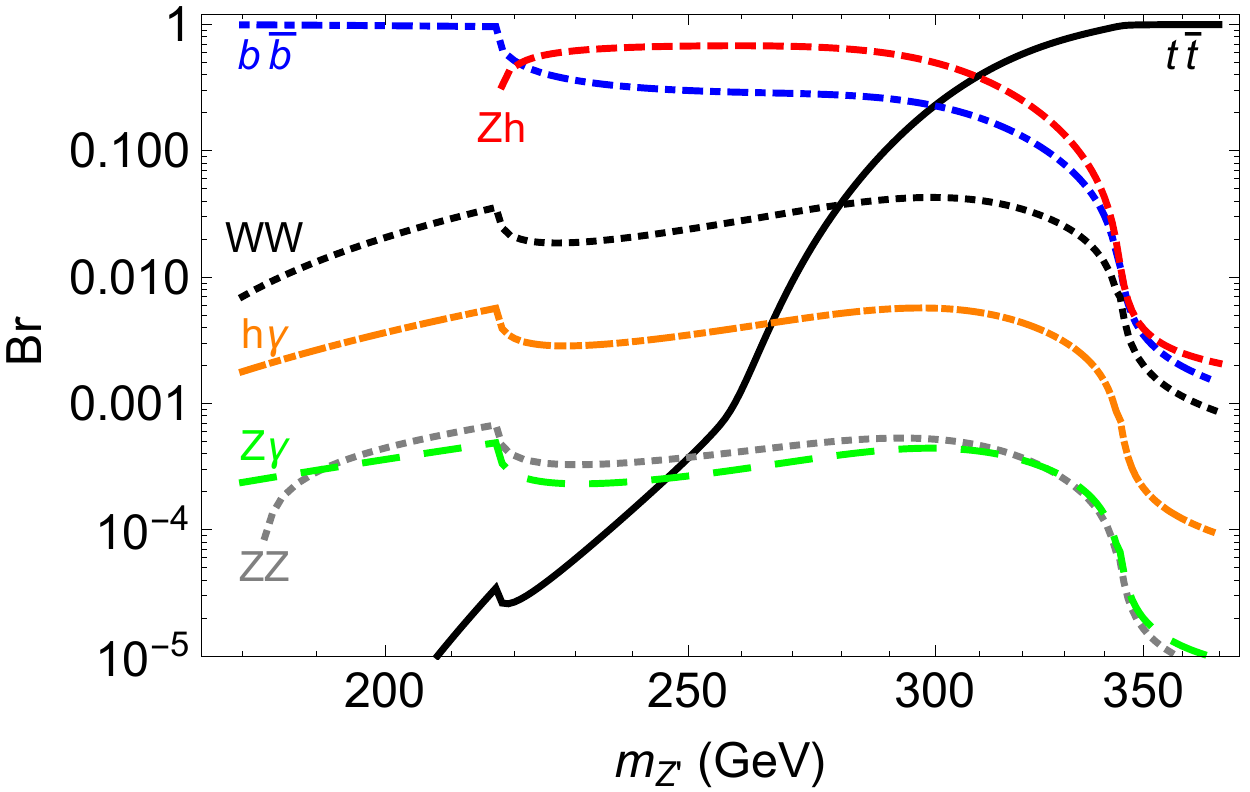}
  \caption{$Z'$ branching ratios as a function of mass, for zero kinetic mixing.} \label{fig:ZpBrnomix}
\end{figure}

\begin{figure}
  \centering
  \includegraphics[width=0.48\textwidth]{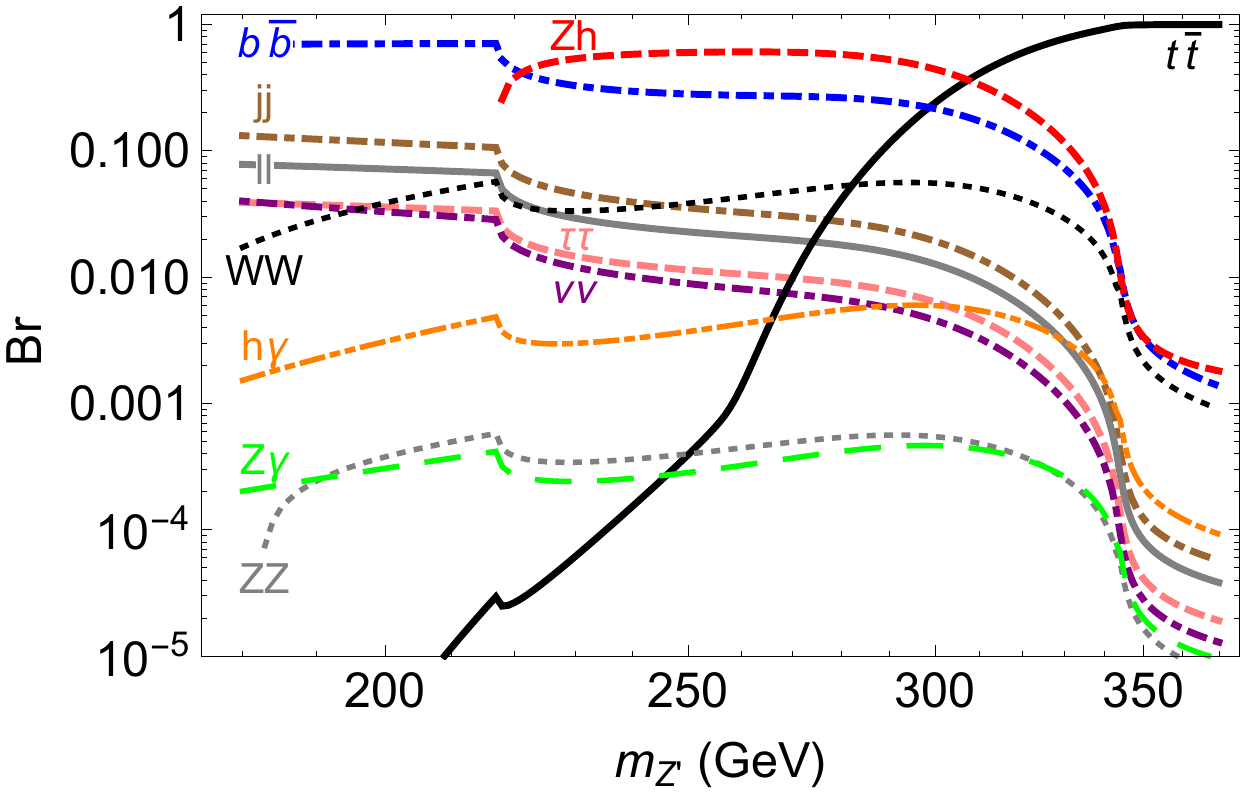}
  \includegraphics[width=0.48\textwidth]{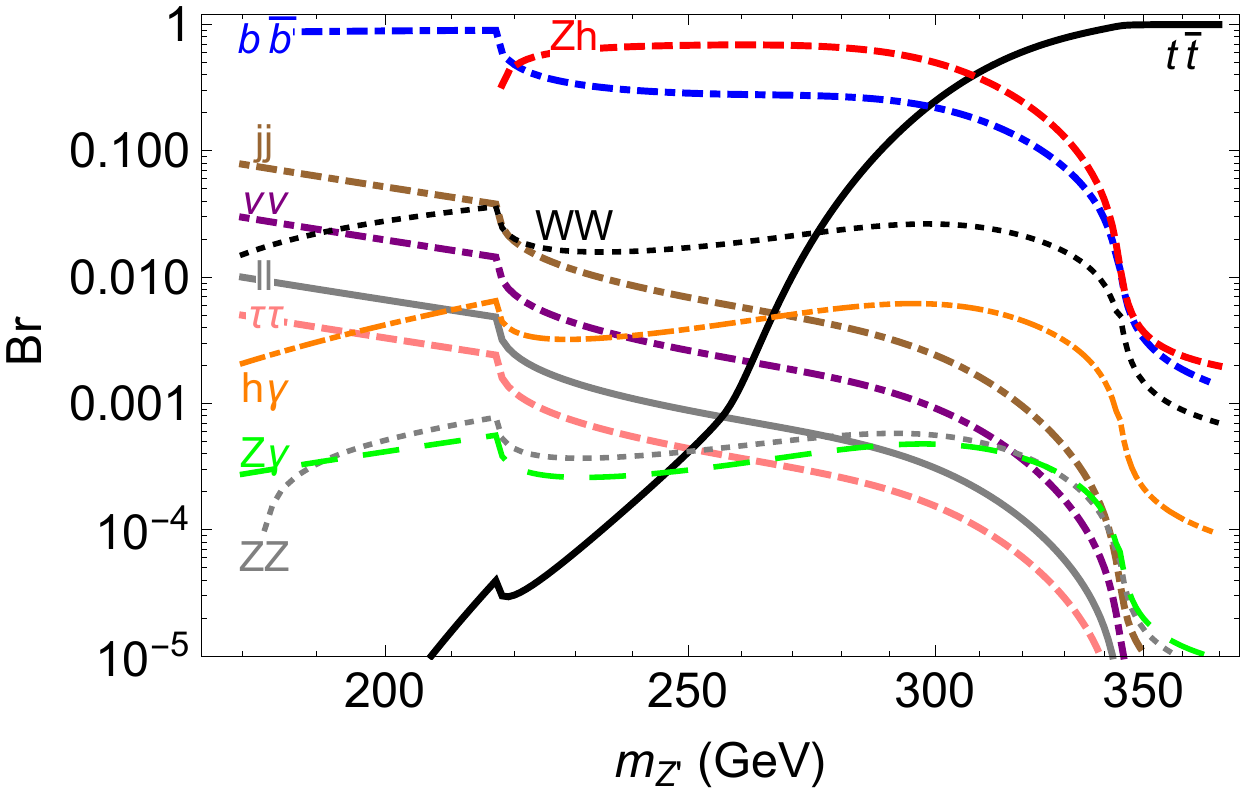}
  \caption{$Z'$ branching ratios as a function of mass, for zero mass mixing and kinetic mixing $\epsilon = 10^{-2}$ (left); and for zero kinetic mixing and maximal mass mixing (right). The decay mode marked as $ll$ is the sum of the $ee$ and $\mu\mu$ decay modes, which are equal up to small kinematic corrections.}\label{fig:ZpBrmix}
\end{figure}

We have computed the decay widths using FeynArts~3.9~\cite{FeynArts} and FormCalc~8~\cite{FormCalc,FormCalcMath}. The branching ratios at zero mixing are shown in \figref{fig:ZpBrnomix}, and for non-zero kinetic (mass) mixing in the left (right) of \figref{fig:ZpBrmix}; analytic expressions are given in \appref{app:dwidth}.  We take the kinetic mixing $\kmix = 10^{-2}$, and the mass mixing to be negative and saturate the bound of \modeqref{eq:dm2limit}.  These sign choices maximise the branching ratio to light quarks and leptons. In all cases, we see that the dominant modes are some combination of $b\bar{b}$, $Zh$ or $t\bar{t}^{(\ast)}$, depending on which channels are kinematically accessible. Decays to $WW$ are a non-trivial fraction, while decays to $Z\gamma$ and $h\gamma$ could offer a potential signal in the form of a hard photon.

\begin{figure}
  \centering
  \includegraphics[width=0.49\textwidth]{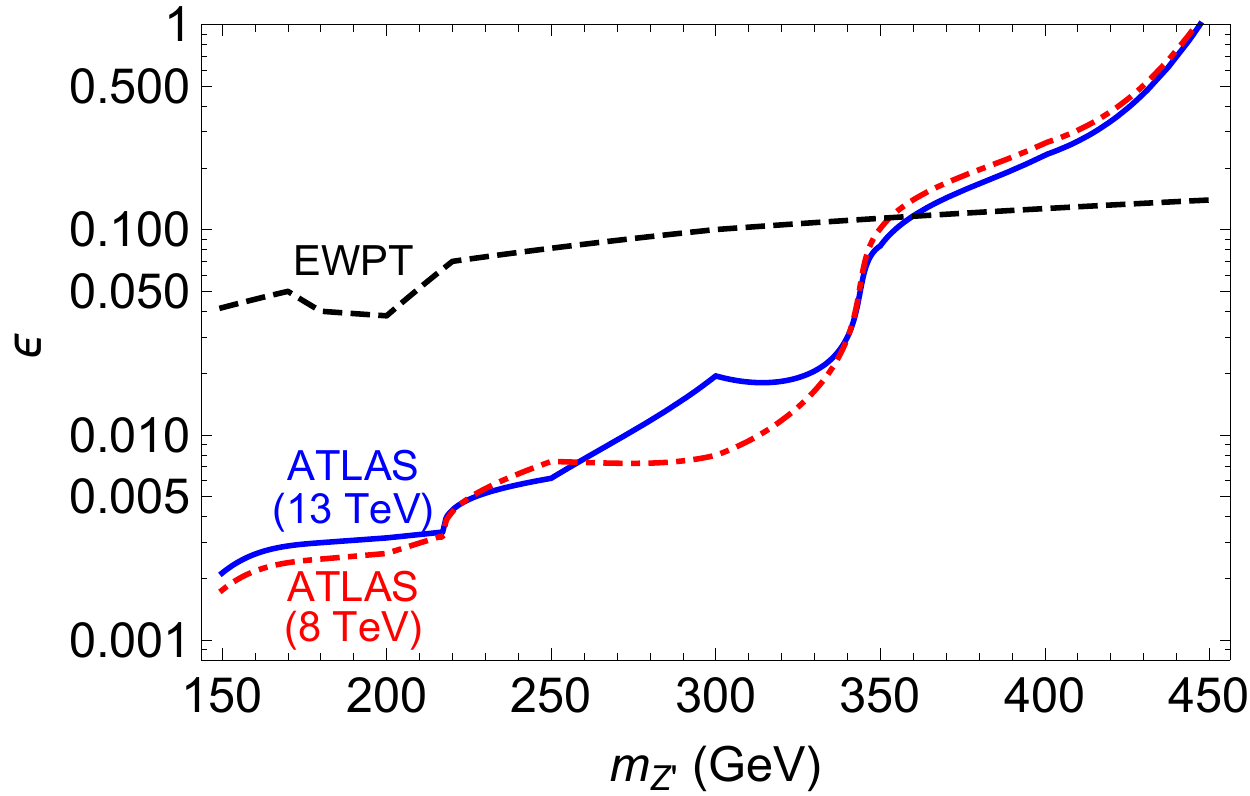}
  \includegraphics[width=0.46\textwidth]{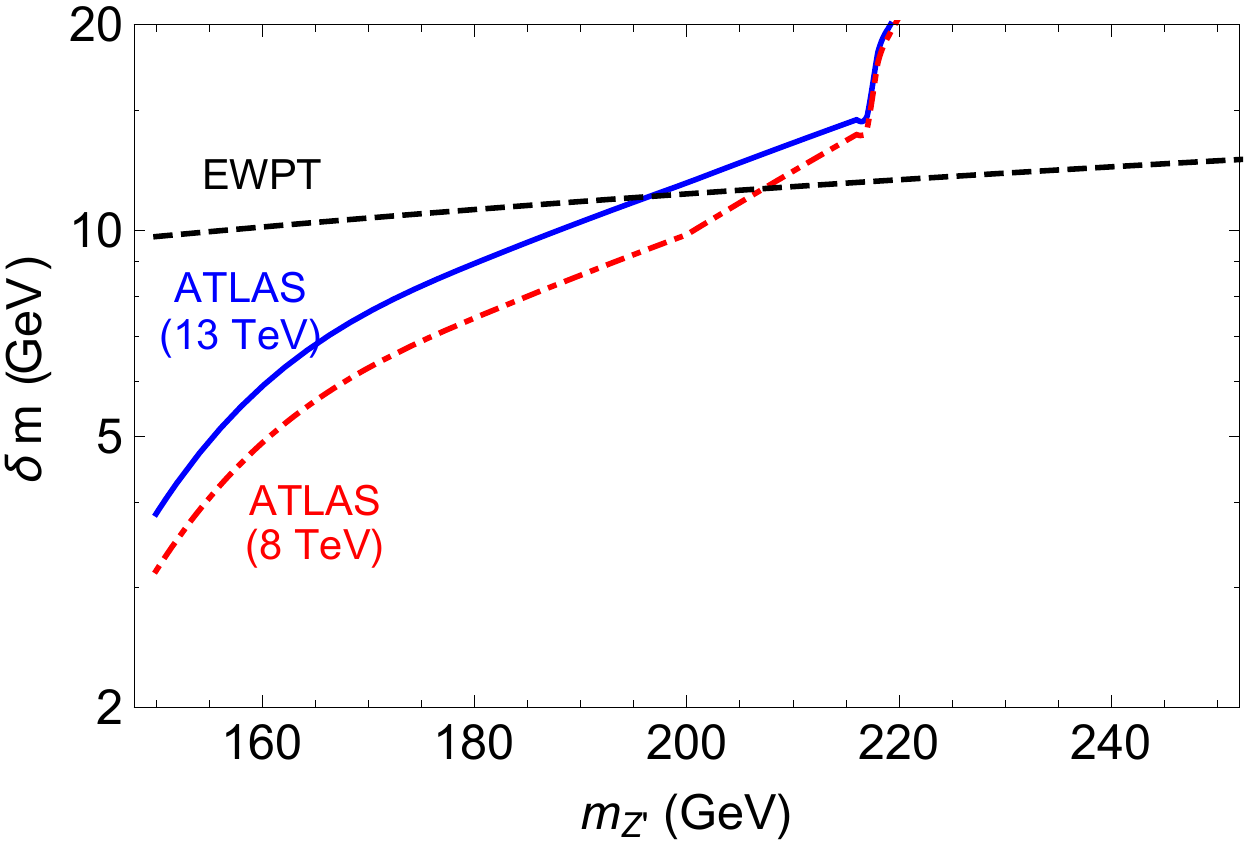}
  \caption{Limits on the kinetic (left) and mass (right) mixing from ATLAS searches for dilepton resonances at 8~\cite{Aad:2014cka} and 13~TeV~\cite{ATLAS-CONF-2015-070}, compared to EWPT bounds~\cite{1006.0973}.}\label{fig:leptonlims}
\end{figure}

When the mixing is non-zero, we have additional decays to electrons and muons that might provide interesting signals.  These branching ratios are larger for kinetic mixing, and decrease more rapidly with mass for mass mixing, because $\kmix$ appears unsuppressed in the matrix elements while the mass mixing is suppressed by $m_{Z'}^2$.  This means that the relative contribution to the decays from mass mixing is proportional to
\begin{equation}
  \frac{\amp^{mass-mixing}}{\amp^{kinetic-mixing}} \sim \frac{\dm}{\kmix m_{Z'}^2} = 0.7 \, \frac{m_Z}{m_{Z'}}\,,
\end{equation}
where in the last equality we have replaced $\kmix$ and $\dm$ by their maximal values.  These decays will lead to additional limits on $Z'$ and $\chi$, but the parameters controlling these decays are irrelevant for the Higgs and DM phenomenology we are interested in.  We show these limits using ATLAS searches for dilepton resonances at 8 and 13~TeV in \figref{fig:leptonlims}.  We see that below the $t\bar{t}$ threshold, direct searches are about an order of magnitude more constraining on kinetic mixing than EWPT, imposing $\epsilon \lesssim 10^{-3}$.  In constrast direct searches only place limits on mass mixing below $m_{Z'} \approx 200$~GeV, and for the range of masses of interest are only moderately better than bounds from $m_Z$.

We will henceforth assume that the kinetic and mass mixings are sufficiently small that the branching ratios are described by \figref{fig:ZpBrnomix}.  Based on these results, we can divide the parameter space into three regions based on the dominant decay mode of the $Z'$. When $m_{Z'} \gtrsim 300$~GeV, the decay $Z' \to \bar{t}t^{(\ast)}$ dominates. The phenomenology here is likely to be similar to that studied in~\cite{1410.6099}. However, below the top threshold several other decay modes can have a sizeable branching fraction. In the intermediate region 220~GeV~$\lesssim m_{Z'} \lesssim 300$~GeV, the decay $Z' \to Zh$ dominates with $Z' \to \bar{b}b$ also significant.  Finally, for $m_{Z'} \lesssim 220$~GeV, the decay to $\bar{b}b$ is most important.

Last we briefly comment on the case $m_{Z'} > 2m_\chi$.  When $2m_t > m_{Z'}$, decays to the dark matter will dominate as all other branching ratios are suppressed by loop factors $g^4(4\pi)^{-2} \sim 10^{-3}$.  The $Z'$ appears as missing energy at the LHC.  When $m_{Z'} \gg 2m_t$, the decay to tops and to DM are comparable for $g_t = g_\chi$.  The width to tops is enhanced by a colour factor of 3, but suppressed by a factor of 2 due to the chiral coupling.  Closer to threshold, kinematic effects modify this ratio.

\subsection{Flavour Constraints}

Finally we make a few comments on the flavour implications of the $Z'$.  If we assume that the $Z'$ couples only to top quarks in the gauge basis, then after rotating to the mass basis it will in general mediate flavour violation.  However, because the $Z'$ only couples to right-handed quarks, the resultant constraints can be trivially satified.  Specifically, let us write the up-type Yukawa couplings in the gauge basis $Y_u$ as
\begin{equation}
  Y_u = U_{L} \, y_u \, U_R^\dagger \,,
\end{equation}
where $y_u$ is the (diagonal) mass-basis Yukawa, and $U_L$ and $U_R$ are unitary matrices corresponding to the transformations from the gauge to mass bases of the left and right-handed quarks respectively.  The flavour violation induced by the $Z'$ is proportional to the off-diagonal elements of $U_R$ (specifically to $U_R^{13}$ and $U_R^{23}$).  However, as is well known in the SM the matrix $U_R$ is unphysical and cannot be measured.  In particular, there is no theoretical problem with taking $U_R^{13} = U_R^{23} = 0$ so that the $Z'$ still only couples to the top in the mass basis.

We briefly comment on how small the mixing parameters must be.  The strongest constraints come from charm physics.  Integrating out the $Z'$ leads to the low-energy effective operator
\begin{equation}
  \lag \supset \frac{g_t^2 \abs{U_R^{13}}^2 \abs{U_R^{23}}^2}{m_{Z'}^2} \, \bigl( \bar{c}_R \gamma^\mu u_R)^2 \,.
\end{equation}
The scale of this operator is constrained to lie above $\gtrsim 10^3$~TeV~\cite{Bona:2007vi,Isidori:2010kg}.  This gives us the approximate bound
\begin{equation}
  \abs{U_R^{13}} \abs{U_R^{23}} \lesssim 3 \times 10^{-4} \, \biggl( \frac{m_{Z'}/g_t}{300 \text{ GeV}} \biggr) \,.
\end{equation}
Bounds on the elements of $U_R$ individually come from rare top decays, and are completely negligible.  We are mainly interested in the region of parameter space where $m_{Z'} > m_t$, so $Z'$ mediates three-body decays, primarily $t \to (u/c) b \bar{b}$.  We can estimate the branching ratio as
\begin{equation}
  \frac{\Gamma(t \to u^i b \bar{b})}{\Gamma_t} \sim \frac{g_t^2 \abs{U_R^{i3}}^2}{(4\pi)^3 \alpha} \, \frac{\Gamma_{Z'}^2 m_W^2}{m_{Z'}^4} \sim \abs{U_R^{i3}}^2 \times 10^{-12} \,.
\end{equation}
This is well below any feasible experimental measurement for the forseable future~\cite{Bardhan:2016txk}.

\section{Collider searches: constraints and improved fits} \label{sec:collider_searches}

In this section we demonstrate that an improved fit to the $\bar{t}th$ data from the LHC can be obtained from our model.  We also discuss current limits and the projected reach of the LHC at $\sqrt{s}=13\,$TeV in the relevant region of the parameter space.

\subsection{Fit to the $\mathbf{t\bar{t}H}$ data} \label{sec:ttH}

The ATLAS and CMS experiments recently presented their combined measurements of the Higgs properties using the complete Run I dataset~\cite{ATLAS-CONF-2015-044,CMS:2015kwa}. While overall the results show good agreement with the Standard Model predictions, there is a mild $2.3\sigma$ excess in the measured signal strength for production of the Higgs in association with top quarks. A more precise measurement of this production mechanism will be one of the key objectives in Run II and it is therefore interesting to consider the possibility that new physics could be responsible for the currently observed excess. We shall consider this possibility in the context of our $Z'$ model. 

In Table~\ref{tab:ttH_results} we have listed the measured ATLAS~\cite{1409.3122,1503.05066,1506.05988} and CMS~\cite{1502.02485,1408.1682} signal strengths in the separate final state channels. It is clear that the dominant contribution to the excess arises from the CMS same-sign dilepton (and in particular the di-muon) channel. In the final column we have combined the measurements using the procedure adopted by the Particle Data Group~\cite{Agashe:2014kda}. In particular we assume that all errors are Gaussian and uncorrelated between the experiments. 

\begin{savenotes}
\begin{table}[h]
  \centering
  \begin{tabular}{|c|c|c|c|}
    \hline
    & ATLAS & CMS & Combined \\
    \hline
    $\gamma\gamma$ & $1.4^{+2.6}_{-1.7}$ & $2.7^{+2.6}_{-1.8}$ & $2.1\pm1.5$ \\
    \hline
    $b\bar{b}$ & $1.5^{+1.1}_{-1.1}$ & $1.2^{+1.6}_{-1.5}$ & $1.4\pm0.9$ \\
    \hline
    $\tau_\mathrm{had}\tau_\mathrm{had}$ & $-9.6^{+9.6}_{-9.7}$ & $-1.3^{+6.3}_{-5.5}$ & $-3.5\pm4.9$ \\
    \hline
    SS dilepton\footnote{Unlike the CMS analysis, the ATLAS analysis imposes a hadronic tau veto. They then consider an additional SS 2l+$\tau_\mathrm{had}$ channel and measure a signal strength of $-0.9^{+3.1}_{-2.0}$.} & $2.8^{+2.1}_{-1.9}$ & $5.3^{+2.1}_{-1.8}$ & $4.2\pm1.4$ \\
    \hline
    3 lepton & $2.8^{+2.2}_{-1.8}$ & $3.1^{+2.4}_{-2.0}$ & $2.4\pm1.5$ \\
    \hline
    4 lepton\footnote{The lower uncertainties in the 4 lepton channel have been truncated by the requirement that the expected signal+background event yield not be negative. We therefore use the upper uncertainty for weighting the measurements in the combination.} & $1.8^{+6.9}_{-2.0}$ & $-4.7^{+5.0}_{-1.3}$ & $-2.5\pm4.1$ \\
    \hline
  \end{tabular}
  \caption{ATLAS and CMS signal-strength measurements for $t\bar{t}h$ production in the various final-state channels.} \label{tab:ttH_results}
\end{table}
\end{savenotes}

When investigating the possible $Z'$ contribution to the various channels we must consider the fact that the signal acceptance $\times$ efficiency is generally expected to differ in our case due to the fact that we have both a spin-1 resonance and potentially different decay modes contributing to the final state. However due to the currently limited statistics, particularly in the leptonic channels, in many cases the analyses do not make extensive use of the detailed kinematics of the final state. We therefore assume that the signal efficiencies are unchanged for our $Z'$ signal as compared to the Higgs. A notable exception is the $b\bar{b}$ channel where the experiments make use of multivariate techniques and the matrix element method to distinguish between the $t\bar{t}h$ signal and dominant $t\bar{t}$+jets background. Such multivariate analyses are challenging to replicate and a complete recasting of the experimental analyses is beyond the scope of this paper. Given that these analyses are optimised to separate the $t\bar{t}h$ signal containing a resonant $b\bar{b}$ pair from the $t\bar{t}+b\bar{b}$ background, we instead make the reasonable assumption that only final states containing a resonant $b\bar{b}$ pair, either from the decay of the $Z'$ or the Higgs (via $Z'{\rightarrow}Zh$), contribute to the signal. We expect this to lead to a stronger limit on the $t\bar{t}Z'$ cross-section than would be obtained by the full analysis. However, our approximation is aided by fact that $Z'{\rightarrow}b\bar{b}$ decays are most important for smaller $Z'$ masses not too far from the Higgs mass, as shown in \figref{fig:ZpBrnomix}. 

\begin{figure}[h]
  \centering
  \begin{minipage}[b]{0.45\textwidth}
    \centering
    \includegraphics[width=0.9\textwidth]{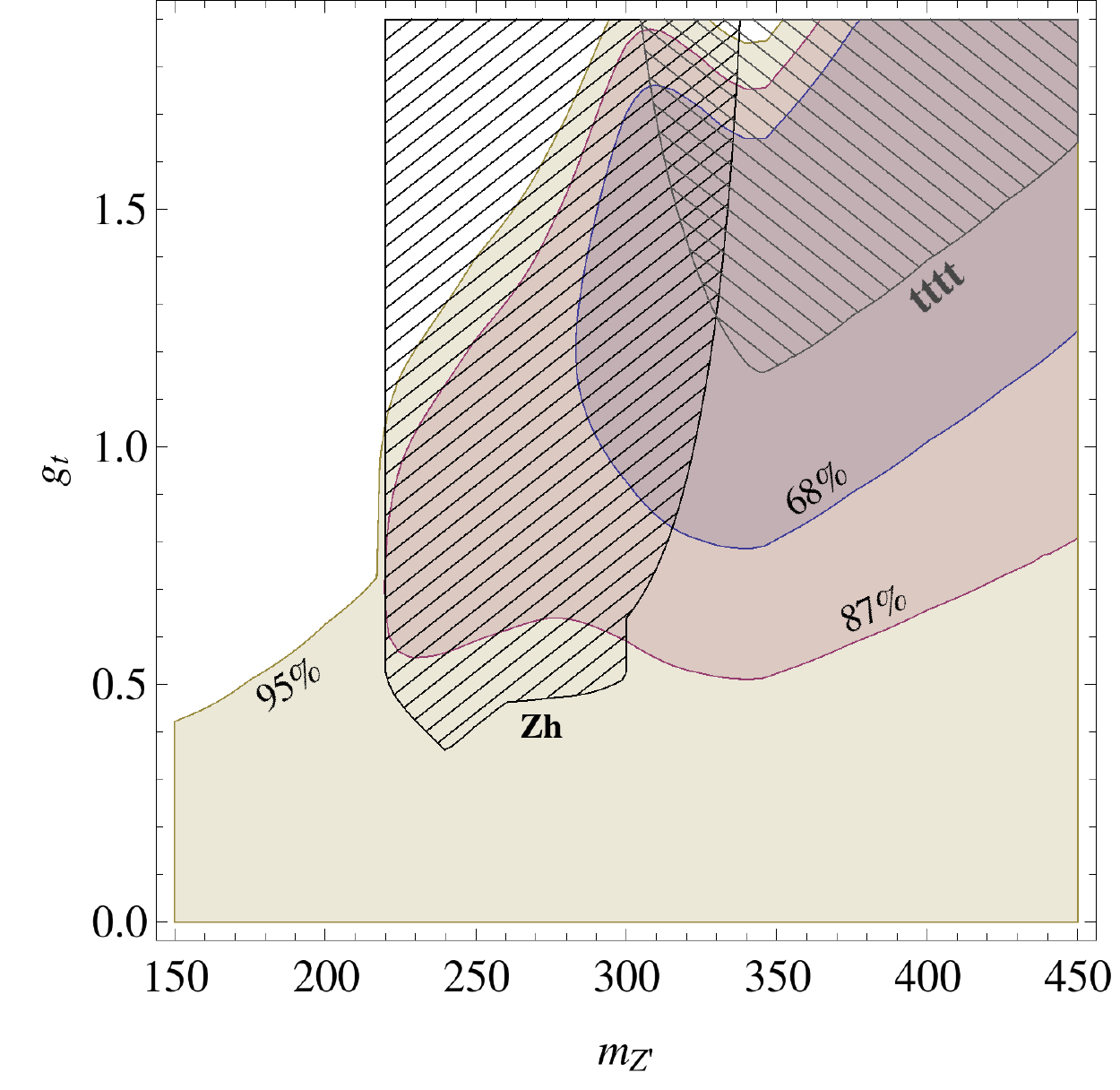}
  \end{minipage}
  \begin{minipage}[b]{0.45\textwidth}
    \centering
    \includegraphics[width=0.9\textwidth]{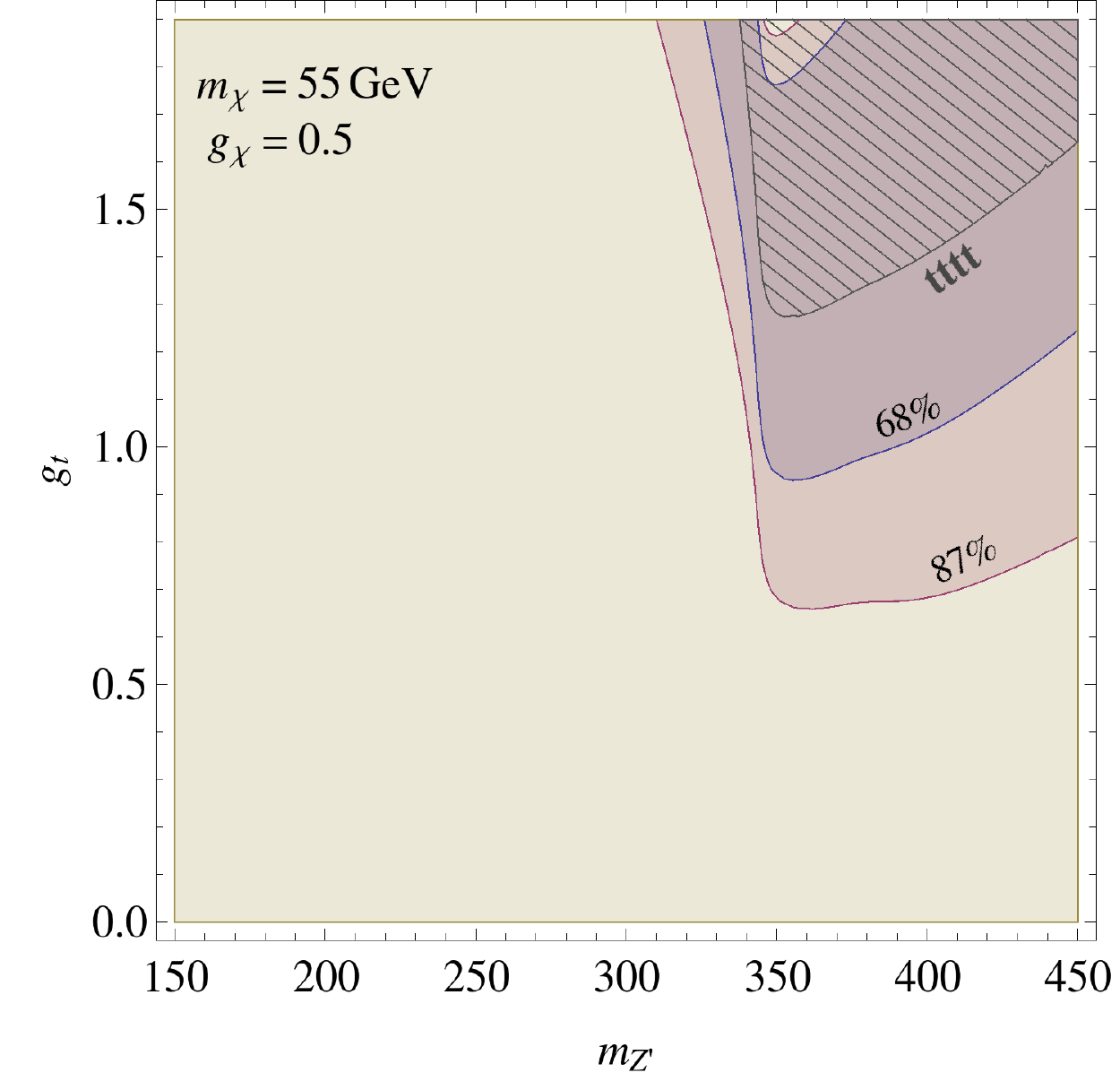}
  \end{minipage}
  \caption{The best-fit region to the measured $t\bar{t}h$ signal strengths in the $m_{Z'}$-$g_t$ plane. The blue, red and yellow shaded regions correspond to the 68\%, 87\% and 95\% confidence levels respectively. The grey and black hatched regions show the limits from $t\bar{t}t\bar{t}$ and $Zh$ searches. The left panel assumes $BR(Z'\rightarrow\mathrm{invisible})=0$, while in the right panel we have taken $m_\chi=55$~GeV and $g_\chi=0.5$.} \label{fig:ttH_fit}
\end{figure}

We perform a $\chi^2$-fit of the signal-strengths listed in Table~\ref{tab:ttH_results} as a function of the $Z'$ mass and coupling, $g_t$. The results are shown in \figref{fig:ttH_fit} where we have plotted the best-fit region at the 68\%, 87\% and 95\% confidence levels. We find that a $Z'$ contribution can indeed explain the observed excess in the $t\bar{t}h$ signal strength, in particular in the high-mass region where the $Z'$ is decaying dominantly into $t\bar{t}^{(*)}$.

\subsection{Current and Projected Limits} \label{sec:LHC_constraints}

There are additional LHC searches which also constrain the parameter space of our $Z'$ model. The relevant searches are largely determined by the dominant branching ratios and can be divided into three mass regions. In the high mass region ($m_{Z'}\gtrsim330$~GeV) the most stringent constraints come from limits on $t\bar{t}t\bar{t}$ production where the $Z'$ is produced in association with top quarks. The upper bound on the cross-section from the ATLAS same-sign dilepton search~\cite{1504.04605} is 70~fb assuming SM kinematics. In order to determine the bound on $t\bar{t}Z'$ production we have simulated $5\times10^4$ events (using {\tt MadGraph5\_aMC@NLO}, {\tt Pythia~6.4}~\cite{Sjostrand:2006za} and {\tt Delphes~3}~\cite{deFavereau:2013fsa}) for a range of $Z'$ masses and calculated a relative signal efficiency with respect to the SM. This factor varies from $\sim0.1-1.2$ across the 5 signal regions in the ATLAS analysis and the $Z'$ masses we considered. For each $Z'$ mass we then rescale the reported limit by the largest efficiency factor across the signal regions with good sensitivity. This is expected to give a slight overestimate compared to the actual limit, although the excluded region is relatively insensitive to the precise efficiencies. 

Intriguingly, this analysis also observed a $2.5\sigma$ excess in the signal regions sensitive to four top quark production. In the case of $t\bar{t}t\bar{t}$ production with SM-like kinematics such a signal is ruled out by the lepton+jets analysis~\cite{1505.04306}, which sees no corresponding excess and gives a stronger upper bound of 23~fb on the cross-section. However this does not necessarily hold in the context of our $Z'$ model where the bound on the cross-section from this search is expected to be weaker. This is due to the fact that $t\bar{t}Z'$ production gives rise to a softer $H_T$ spectrum than the SM case for the $Z'$ mass range we are considering. Unfortunately determining the precise bound on $t\bar{t}Z'$ production is not straightforward since the experimental analysis performs a fit to the $H_T$ spectrum across eight signal regions. It would certainly be interesting for the experiments to also consider a signal topology similar to our $Z'$ scenario in the future.

In the intermediate mass region ($220\,\,{\lesssim}\,\,m_{Z'}\lesssim330$~GeV) the strongest constraints arise from $Z'$ decays into $Zh$. ATLAS has performed a search~\cite{1503.08089} for a massive vector decaying into $Zh$, which provides constraints on loop-induced $Z'$~+~jets production. This analysis only considers resonance masses above 300~GeV, so for lower $Z'$ masses we use the CP-odd Higgs search~\cite{1502.04478}, which gives very similar bounds in the regions where they overlap. In order to estimate the signal acceptance $\times$ efficiency for the case of a spin-1 resonance we have rescaled the bound from Ref.~\cite{1502.04478} by an additional factor of $\sim1.5$ corresponding to the ratio of the expected limits for the two searches at a mass of 300~GeV.

For the low mass region ($m_{Z'}\lesssim220$~GeV) the $t\bar{t}h$ searches discussed in the previous section currently provide the only relevant LHC constraints. In this region the $Z'$ is decaying dominantly into $b\bar{b}$ and searches targeting the $Z'$~+~jets production mechanism in this decay mode will be impossible at a hadron collider. However other decay modes are likely to provide useful constraints in the future despite their relatively small branching ratios. In particular there are existing $Z\gamma$ resonance searches~\cite{1407.8150} that are sensitive to $Z'$ masses above 200~GeV and which will become relevant at $\sqrt{s}=13-14$~TeV. Another channel which may be able to provide constraints in the low mass region is $Z'{\rightarrow}h\gamma$. It would be interesting to further investigate the potential sensitivity of this channel. Finally, there is also the possibility of limits from LEP searches for $Z'$ masses below $209$~GeV. However the production cross-section is negligible unless one considers non-zero kinetic mixing. The limits on production through kinetic mixing were included in Ref.~\cite{1006.0973}, and away from the $Z$-pole were found to be subdominant to constraints from corrections to $m_Z$. 

In \figref{fig:ttH_fit} we have overlaid the current limits from the $t\bar{t}t\bar{t}$ and $Zh$ searches on the best fit region for the $t\bar{t}h$ excess. As can be seen on the left of \figref{fig:ttH_fit}, which corresponds to the case of only visible decays for the $Z'$, these searches are already starting to probe a significant region of the parameter space including in the higher mass regions which also fit the $t\bar{t}h$ signal. Nevertheless there currently remains a region which both provides a good fit to the $t\bar{t}h$ excess and satisfies all other bounds. When a non-zero branching ratio into DM is kinematically allowed, as shown in the example on the right of \figref{fig:ttH_fit}, we see that the region providing a good fit to the $t\bar{t}h$ signal is reduced. This is due to the significantly suppressed branching ratio to visible final states below the $t\bar{t}$ threshold. The best-fit region also extends to larger values of $g_t$ in order to compensate for the reduced branching ratio. Also notice that the constraints from the $Zh$ resonance search completely vanish, while the four top searches lose their sensitivity as soon as one of the top quarks from the decay becomes off-shell. 

\begin{figure}[h]
  \centering
  \begin{minipage}[b]{0.45\textwidth}
    \centering
    \includegraphics[width=0.9\textwidth]{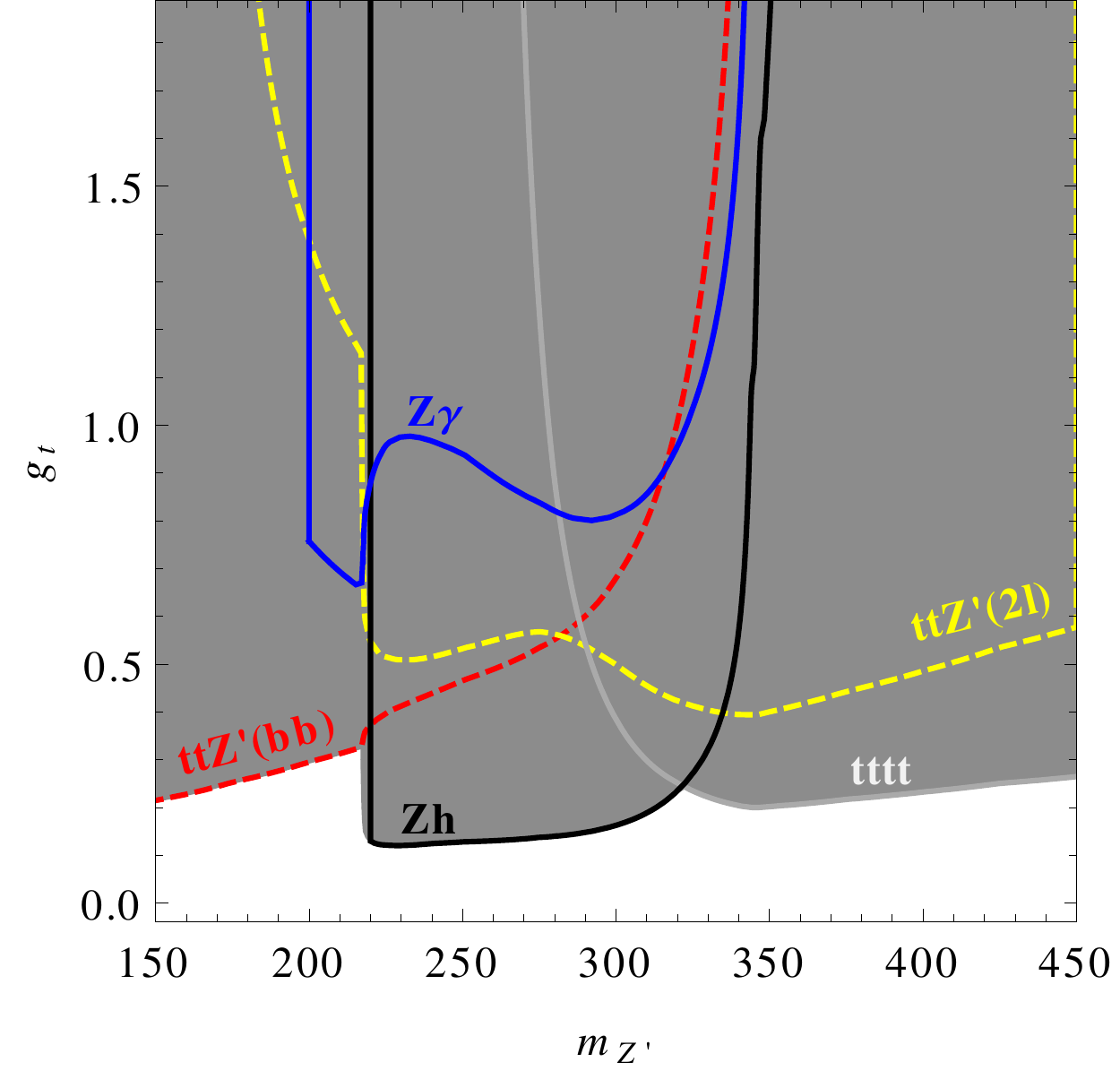}
  \end{minipage}
  \begin{minipage}[b]{0.45\textwidth}
    \centering
    \includegraphics[width=0.9\textwidth]{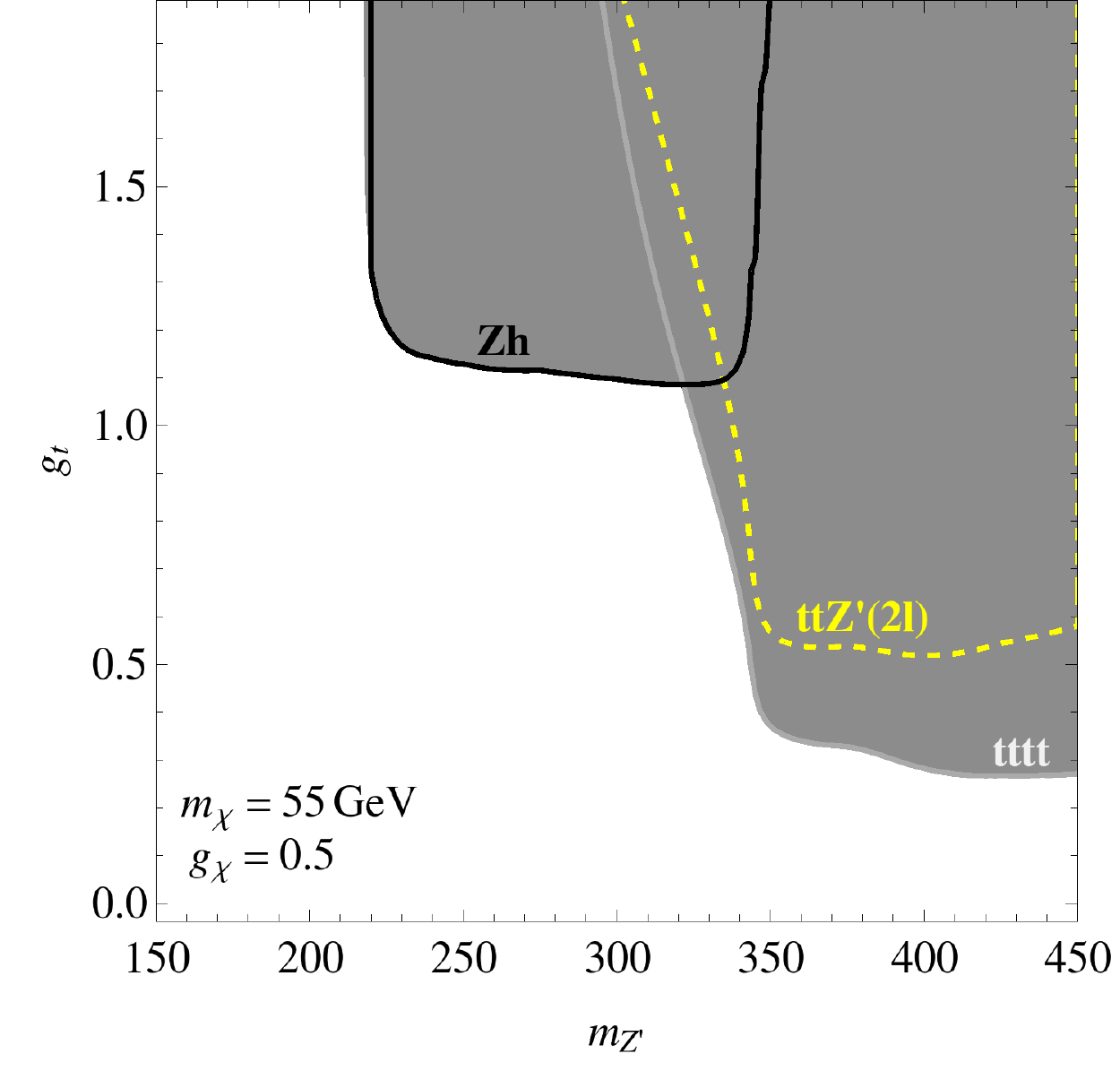}
  \end{minipage}
  \caption{Projected limits from LHC searches at $\sqrt{s}=13$~TeV with $\mathcal{L}=300\,\mathrm{fb}^{-1}$ integrated luminosity. The left panel assumes $BR(Z'\rightarrow\mathrm{invisible})=0$, while in the right panel we have taken $m_\chi=55$~GeV and $g_\chi=0.5$. The black and blue lines correspond to $Zh$ and $Z\gamma$ resonance searches respectively while the $t\bar{t}t\bar{t}$ search is denoted by the grey line. An estimate of the projected sensitivity to $t\bar{t}Z'$ production based on existing $tth$ searches is shown by the red and yellow dashed lines for the $b\bar{b}$ and same-sign dilepton channels respectively.} \label{fig:LHC_constraints}
\end{figure}

We also show in \figref{fig:LHC_constraints} the projected 95\% CLs exclusion for the various searches at $\sqrt{s}=13$~TeV with 300~fb$^{-1}$ of integrated luminosity. The significant improvement in the limits is predominantly driven by the large increases in the $Z'$ cross-sections when moving to the higher center of mass energy. The cross-sections for the dominant backgrounds in the $t\bar{t}t\bar{t}$, $Zh$ and $Z\gamma$ analyses ($t\bar{t}$, $Z+\mathrm{jets}$ and $Z\gamma$) increase by more modest factors of 3.3~\cite{1112.5675}, 1.5~\cite{1208.5967} and 1.7~\cite{1105.0020} respectively. We also include the projected limits from $t\bar{t}h$ searches assuming a Standard Model Higgs and the absence of any continuing excess. Although note that our projections are likely to underestimate the sensitivity for these searches since they are expected to differ substantially in Run II where the increased number of events will allow for greater exploitation of the kinematics and reconstruction of the resonance mass. In deriving the limits we have assumed that the significance is simply given by $S/\sqrt{B}$ such that the limits scale with $\sqrt{\mathcal{L}}$ after taking into account the relative increases in the signal and background cross-sections. In the intermediate and high-mass regions there are good prospects for limits on both $Z'$~+~jets and $t\bar{t}Z'$ production, while the latter will provide the only significant constraints in the low mass region where the $Z'$ is decaying dominantly to $b\bar{b}$. Once again, notice that when invisible decays are kinematically allowed, as shown on the right of \figref{fig:LHC_constraints}, the constraints from future searches are drastically ameliorated below the $t\bar{t}$ threshold.

\section{Dark Matter and galactic $\gamma$-ray excess} \label{sec:dmge}

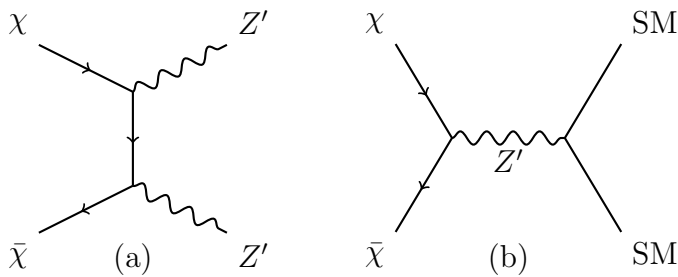
\begin{figure}
  \centering
  \parbox{0.3\textwidth}
  {
    \centering
    \begin{tikzpicture}[node distance=0.625cm and 1.25cm]
      \coordinate (v1);
      \coordinate[above left = of v1, label=above left:{$\chi$}] (i1);
      \coordinate[above right = of v1, label=above right:{$Z'$}] (o1);
      \coordinate[below = of v1] (vb);
      \coordinate[below = of vb] (v2);
      \coordinate[below left = of v2, label=below left:{$\bar{\chi}$}] (i2);
      \coordinate[below right = of v2, label=below right:{$Z'$}] (o2);
      \coordinate[below = of v2, label=below:{(a)}] (va);
      \draw[fermion] (i1) -- (v1);
      \draw[fermion] (v1) -- (v2);
      \draw[fermion] (v2) -- (i2);
      \draw[photon] (v1) -- (o1);
      \draw[photon] (v2) -- (o2);
    \end{tikzpicture}
  }
  \parbox{0.3\textwidth}
  {
    \centering
    \begin{tikzpicture}[node distance=1.25cm and 0.75cm]
      \coordinate (v1);
      \coordinate[above left = of v1, label=above left:{$\chi$}] (i1);
      \coordinate[below left = of v1, label=below left:{$\bar{\chi}$}] (i2);
      \coordinate[right = of v1, label=below:{$Z'$}] (va);
      \coordinate[right = of va] (v2);
      \coordinate[above right = of v2, label=above right:{SM}] (o1);
      \coordinate[below right = of v2, label=below right:{SM}] (o2);
      \coordinate[below = of va, label=below:{(b)}] (vb);
      \draw[fermion] (i1) -- (v1);
      \draw[fermion] (v1) -- (i2);
      \draw[photon] (v1) -- (v2);
      \draw[fermionnoarrow] (o1) -- (v2) -- (o2);
    \end{tikzpicture}
  }
  \caption{Dark matter annihilation channels for relic density and cosmic ray signals.  (a): $t$-channel annihilation to two $Z'$, relevant when $m_\chi > m_{Z'}$ (there is also a $u$-channel diagram). (b): $s$-channel annihilation through on- or off-shell $Z'$ to SM.}\label{fig:DMann}
\end{figure}

The Lagrangian given in \modeqref{eq:lagrangian} incorporates a natural Dirac DM candidate $\chi$, which couples only to our $Z'$ gauge boson with a coupling strength $g_{\chi}$, in principle unrelated to $g_{t}$. Depending on the mass of the DM, $m_{\chi}$, there can be several channels that dominate its annihilation. If $m_{\chi} > m_{Z'}$ then there is a t/u-channel annihilation involving the exchange of the DM particle $\chi$ as shown in \figref{fig:DMann}\,(a). The s-wave part of the cross-section can be written as
\begin{equation}
\langle\sigma_{\chi\bar{\chi}\to Z'Z'}v\rangle\approx 
\frac{g_{\chi}^4\left(1-\frac{m^2_{Z'}}{m^2_{\chi}}\right)^{3/2}}{16\pi m^2_{\chi}\left(1-\frac{m^2_{Z'}}{2m^2_{\chi}}\right)^2}\,,
\end{equation}
including both the t and u-channel contributions. This channel can be potentially important in a large region of parameter space. In the case that $m_{\chi}<m_{Z'}$, annihilation is dominated by the s-channel exchange of a $Z'$ into SM particles as seen in \figref{fig:DMann}\,(b).  The relative contributions of different final states is well approximated by the $Z'$ branching ratios for $m_{Z'} = 2m_\chi$ (see \figsref{fig:ZpBrnomix} and \ref{fig:ZpBrmix}). In the limit of small $Z-Z'$ mixing, as was discussed previously for the $Z'$-branching ratios, there are four channels that dominate the annihilation depending on the DM mass: $t\bar{t}$ for $m_{\chi}\geq m_t$, $tWb$ for $155$ GeV $\lesssim m_{\chi}< m_t$, $Zh$ for $(m_h+m_Z)/2 \leq m_{\chi}\lesssim 155$ GeV and $b\bar{b}$ for $m_{\chi}<(m_{Z}+m_h)/2$. For annihilations into $t\bar{t}$ and $b\bar{b}$, we find that the s-wave part of the cross-section takes the form,
\begin{equation}
\langle\sigma_{\chi\bar{\chi}\to t\bar{t}}v\rangle\approx \frac{3g_{\chi}^2 g_{t}^2(4m^2_{\chi}-m^2_t)\sqrt{m^2_{\chi}-m^2_t}}{8\pi m_{\chi}(m^2_{Z'}-4m^2_{\chi})^2}\,,\qquad \langle\sigma_{\chi\bar{\chi}\to b\bar{b}}v\rangle\approx \frac{3g_{\chi}^2 g_{b\bar{b},eff}^2(2m^2_{\chi}+m^2_b)\sqrt{m^2_{\chi}-m^2_b}}{2\pi m_{\chi}(m^2_{Z'}-4m^2_{\chi})^2}\,,
\end{equation}
where $g_{b\bar{b},eff}$ is the effective loop-induced coupling.

For annihilations into $Zh$, the s-wave contribution to the annihilation cross-section is given by
\begin{equation}
  \langle\sigma_{\chi\bar{\chi}\to Zh}v\rangle \approx \frac{g_{\chi}^2g_{Zh,eff}^2} {1024\pi m^4_{\chi}m^2_{Z}} \, \frac{I(4m_\chi^2, m_h^2, m_Z^2) + 48 m^2_{\chi}m^2_{Z}}{(m^2_{Z'}-4m^2_{\chi})^2} \sqrt{I(4m_\chi^2, m_h^2, m_Z^2)} \,,
\end{equation}
where $g_{Zh,eff}$ is the effective loop-induced coupling of the $Z'$ to a $Z$-gauge boson and a Higgs, and $I$ is the triangle function
\begin{equation}
  I(a, b, c) = a^2 + b^2 + c^2 - 2ab - 2ac - 2bc \,.
\label{eq:Idef}\end{equation}
 
Including all the main annihilation channel contributions, one can approximately solve the Boltzmann equation for the DM number density and simply relate the DM relic density today to the total s-wave annihilation cross-section at freeze-out, $a_{total}$, as
 \begin{equation}
 \Omega_{\chi}h^2\approx \frac{1.04\times10^{9} \; {\rm GeV}}{M_{Pl}}\frac{x_F}{\sqrt{g_{*}}}\frac{1}{a_{total}}\,,\label{eq:relicdens}
 \end{equation}
where
\begin{equation}
x_F=\log \left(\frac{1}{2}(\frac{1}{2}+2)\sqrt{\frac{45}{8}}\frac{g}{2\pi^3}\frac{m_{\chi}M_{Pl} a_{total}}{\sqrt{g_{*} x_F}}\right)\,.
\end{equation}
Here, $g=4$ is the number of degrees of freedom of the fermionic Dirac DM, $M_{Pl}=1.22\times 10^{19}$ GeV the Planck mass, $x_F=m_{\chi}/T_F$ with $T_F$ the freeze-out DM temperature, and $g_{*}$ the number of relativistic degrees of freedom at $T_F$.

\begin{figure}
  \centering
  \includegraphics[width=0.6\textwidth]{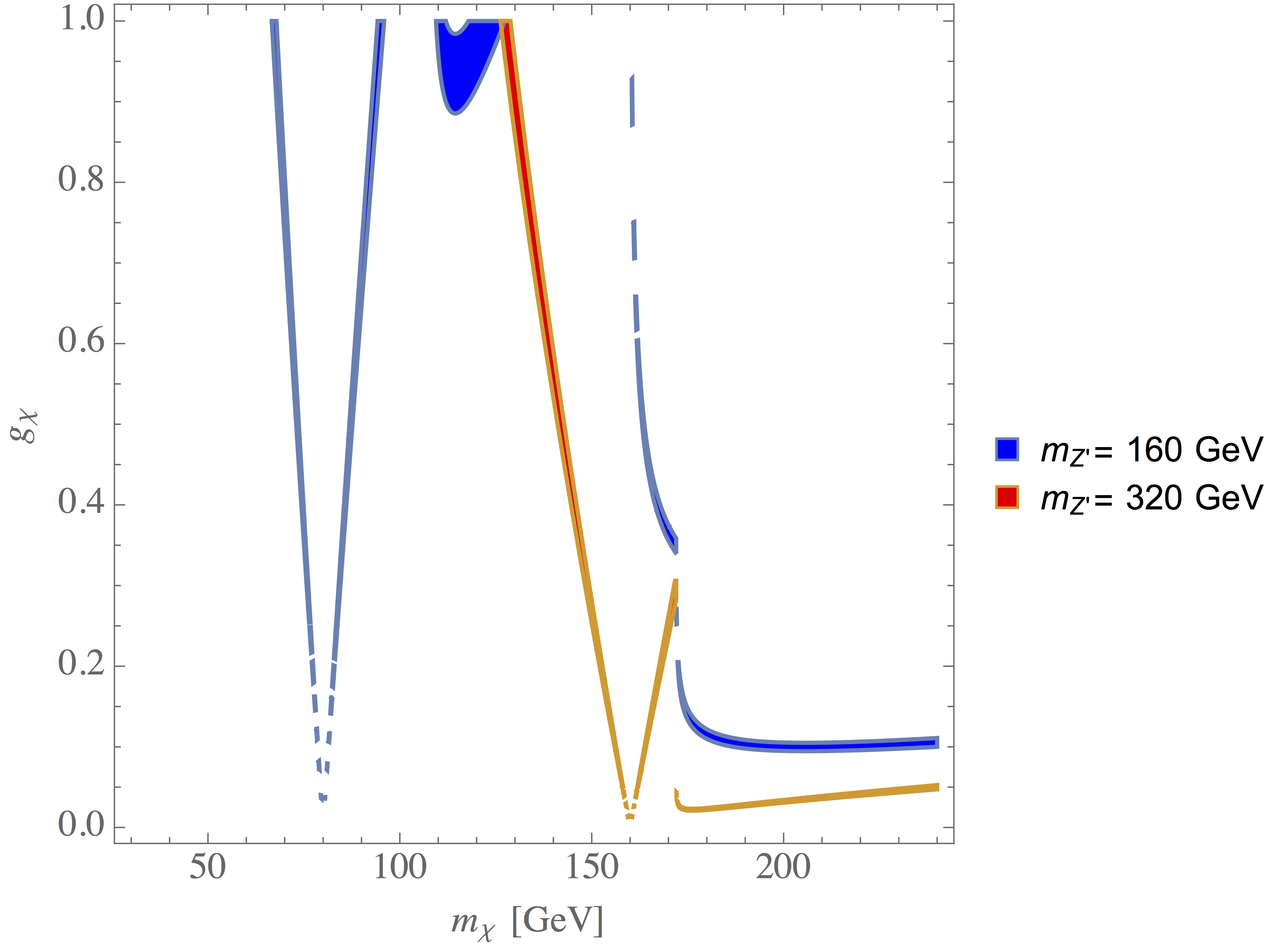}
  \caption{Regions in the ($m_{\chi}$-$g_{\chi}$) plane consistent with the Planck satellite measurement of the DM relic density, considering the main annihilation channels for $\chi$. We have fixed $g_t=0.6$.} \label{fig:Relicdensity}
\end{figure}

We show in \figref{fig:Relicdensity} the regions in the $m_{\chi}$-$g_{\chi}$ plane that are in agreement with the 2$\sigma$-bound on the DM relic density measurement by the Planck satellite experiment, $\Omega_{DM} h^2 \in [0.107,0.131]$~\cite{1502.01582}, except when close to resonance $2m_\chi \approx m_{Z'}$ in which case Eq.~(\ref{eq:relicdens}) breaks down. In the examples shown we have fixed $g_{t}=0.6$, which implies an effective $Z'Zh$ coupling of $g_{Zh,eff}\approx 7$ GeV and an effective $Z'b\bar{b}$ coupling $g_{b\bar{b},eff}\approx 4.5 \times 10^{-3}$, with a very mild $m_{Z'}$ dependence for the values of interest. For $m_{\chi}\geq m_{t}$, annihilations into pairs of top quarks dominate.  The unsuppressed cross section implies a sharp drop in the allowed values of $g_{\chi}$ as $m_{\chi}$ increases across the threshold. For lighter dark matter, there are  two cases to consider.  When annihilations to $Z'$ are possible, they will dominate and set the relic density.  We see in \figref{fig:Relicdensity} for $m_{Z'} = 160$~GeV, the required value of $g_\chi$ increases significantly for $m_\chi < 160$~GeV.  When $m_\chi < m_{Z'}$, all annihilation channels are loop suppressed and the required $g_\chi$ are typically very large.  The exception occurs near the resonance $2m_\chi \approx m_{Z'}$, clearly visible in \figref{fig:Relicdensity} for both vector masses we consider.  The dominant final states are exactly as expected from \figref{fig:ZpBrnomix}; annihilation into $Zh$ comprises 85 percent of the total cross-section until it becomes kinematically unavailable and annihilation into $b\bar{b}$ dominates.

Regarding direct detection experiments, they are mostly insensitive to this kind of model due to the small $Z-Z'$ mixing. The $Z'$ coupling to light quarks (and hence to protons and neutrons) is generated through the kinetic and mass mixings~\cite{Jackson:2013rqp,1402.1173,1411.3342}.  As discussed in \secref{sec:model}, we can always fine-tune these mixings to be small to avoid these constraints. The current LUX constraints~\cite{Akerib:2015rjg} bound the DM-nucleon scattering to $\sigma_n \lesssim 10^{-45}$~cm$^2$ for $m_\chi = 100$~GeV.  For kinetic mixing this gives the bound $g_\chi \, \kmix \lesssim 4 \times 10^{-4}$~\cite{Jackson:2009kg,Jackson:2013rqp}, which requires at most a tuning of $\mathcal{O}(20\%)$. 

There can also be a scattering induced by the $Z'$ coupling to gluons.\footnote{This interaction is non-vanishing despite the Landau-Yang theorem due to the off-shell nature of the $Z'$ in t-channel exchange scattering of the DM against the gluon sea in neutrons and protons.} The effective coupling at zero momentum transfer is
\begin{equation}
  \mathcal{L}_{eff} = \frac{g_t \alpha_s}{24\pi m_t^2} \, \epsilon^{\mu\nu\rho\sigma} F_{\tau\sigma} G_{\nu\rho} G^\tau_{\phantom{\tau}\mu} \,,
\end{equation}
with $G$ ($F$) the gluon ($Z'$) field strength. There is a suppression by $m_t^{-2}$ from the top loop that (in contrast to the effective Higgs-gluon coupling) is not compensated for by a large coupling. We can easily estimate the parametric DM-nucleon scattering cross section:
\begin{equation}
  \sigma \sim \frac{g_t^2 g_\chi^2 \alpha_s^2}{36\pi^3} \, \frac{m_n^4}{m_t^4} \, \frac{\mu_\chi^2}{m_{Z'}^4}  \sim 10^{-47} \text{cm}^2 \, g_t^2 g_\chi^2 \biggl( \frac{100 \text{ GeV}}{m_{Z'}} \biggr)^4 \,.
\end{equation}
Here, $\mu_\chi$ is the nucleon-DM reduced mass. We have assumed $m_\chi \gg m_n$ so $\mu_\chi \sim m_n$. This is well below current experimental limits.

There has recently been excitement about an excess in the $\gamma$-ray flux emanating near the center of our own galaxy and detected by the Fermi-LAT satellite experiment~\cite{1511.02938}. The spectrum of this Galactic Center Excess (GCE) peaks at around 3 GeV and though it could have an astrophysical origin~\cite{Gordon:2013vta,Abazajian:2014fta,Carlson:2014cwa,Yuan:2014rca,Petrovic:2014uda}, interestingly enough it can be well fitted by DM annihilation~\cite{Goodenough:2009gk,Hooper:2013rwa,Calore:2014xka,Gherghetta:2015ysa}. The cross-section that is required to explain the GCE is of the same order as that necessary to account for the DM relic density from thermal freeze-out. The large astrophysical uncertainties and the high-energy tail of the spectrum allow for several annihilation channels to individually provide a good fit~\cite{Calore:2014nla}. This is the case for annihilation into bottom-quark pairs, Higgs-pairs and even top-quark pairs~\cite{Agrawal:2014oha}. The latter does not provide a very good fit in analyses of the GCE using older data from the Fermi-LAT collaboration~\cite{Calore:2014nla}. However, the Fermi collaboration recently reported their own analysis on the excess~\cite{1511.02938}, which seems to confirm preliminary evidence that the GCE can be fit by a somewhat harder spectrum~\cite{MurgiaTalk:2014FermiSymposium,Agrawal:2014oha}.

\begin{figure}
  \centering
  \includegraphics[width=0.45\textwidth]{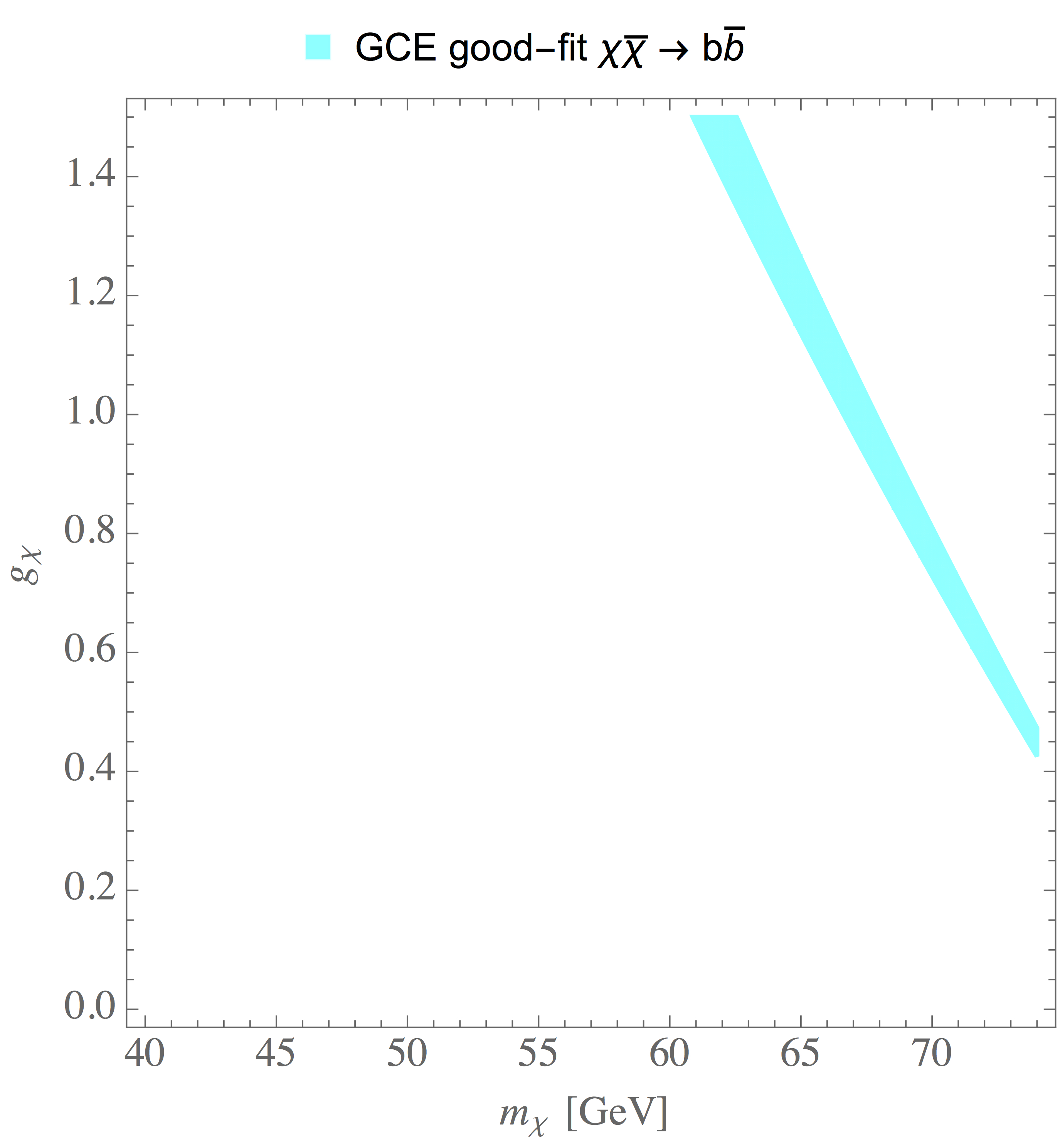}
  \includegraphics[width=0.46\textwidth]{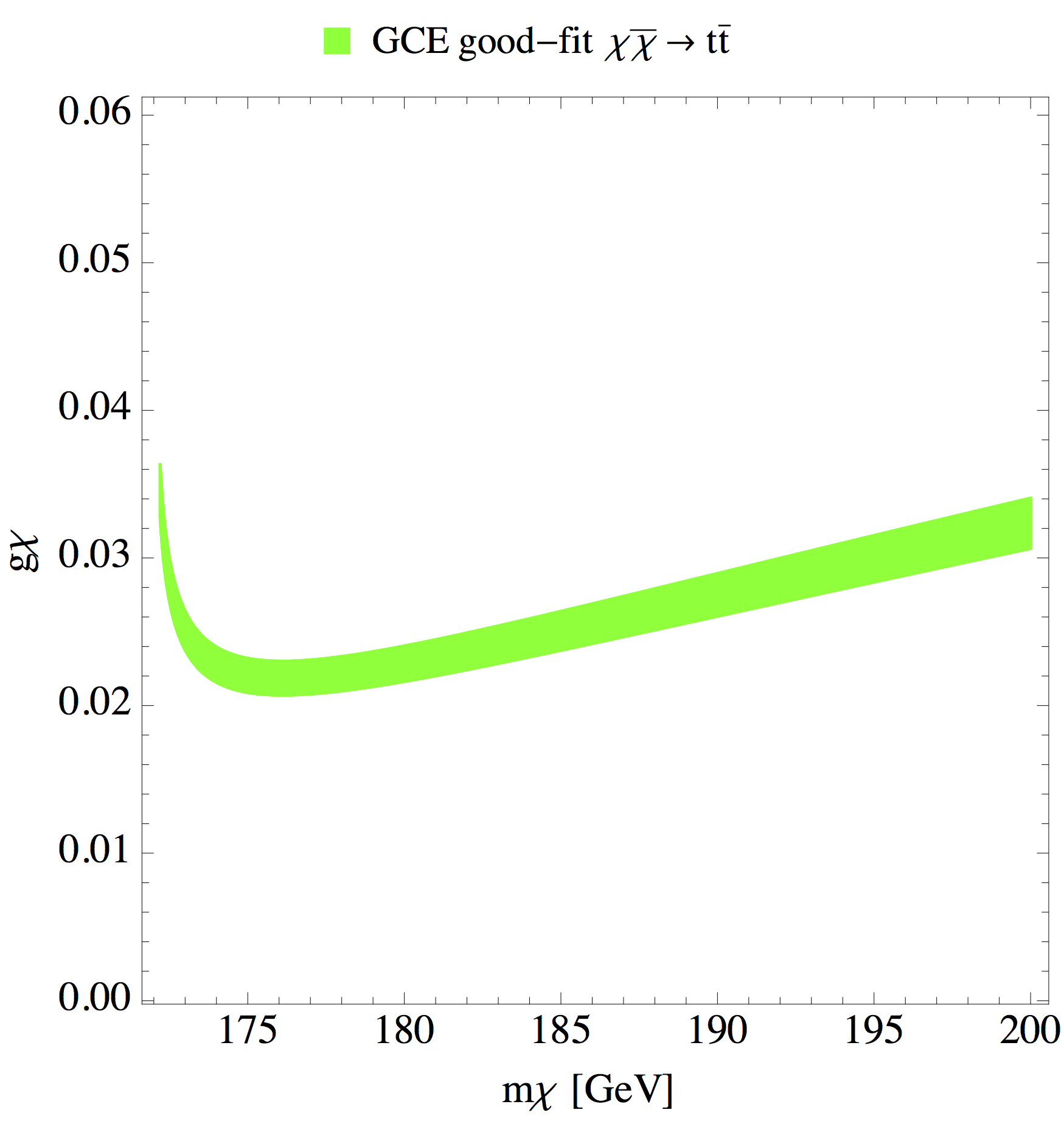}
  \caption{Regions in the ($m_{\chi}$-$g_{\chi}$) plane consistent with the GCE fits for annihilation into $b\bar{b}$ (left) and $t\bar{t}$ (right) performed in Ref.~\cite{Calore:2014nla} and the relic density measurement by Planck. We have taken $g_t=0.6$ and $m_{Z'}=160\,$GeV (left) and $g_t=0.8$ and $m_{Z'}=320\,$GeV (right).} \label{fig:GCE}
\end{figure}

It is therefore interesting to investigate in which regions of parameter space our model could account for the GCE. The fits to the GCE have been performed mainly for annihilations into bottom quarks. Taking into account astrophysical uncertainties, it was found that dark matter masses in the approximate range $m_{\chi}\in [30,74]$ GeV provide a good fit~\cite{Calore:2014nla}. In the case of annihilation into top-quarks, the best fits are accomplished for $m_{\chi}\in [m_t, 200]$ GeV~\cite{Calore:2014nla}. We show in \figref{fig:GCE} the regions of $(g_{\chi},m_{\chi})$ which provide a good fit at 2-$\sigma$ according to the results found in Ref.~\cite{Calore:2014nla}. The left and right panels are for annihilations into $b\bar{b}$ and $t\bar{t}$ respectively. In the $b\bar{b}$ case we have taken $g_{t}=0.6$ and $m_{Z'}=160$ GeV while for $t\bar{t}$ we have fixed $g_{t}=0.8$ and $m_{Z'}=320$ GeV. The former is a point in parameter space that is allowed by current collider constraints (see right panel of Fig.~\ref{fig:ttH_fit}) due in particular to the kinematically allowed invisible decay of the $Z'$, while the latter is a point in the parameter space which also allows for a good fit of the $t\bar{t}h$ excess as shown on the left panel of  Fig.~\ref{fig:ttH_fit} since in this case $Z'$ invisible decays are not kinematically allowed.  The loop-generated mixings discussed in \secref{sec:model} are consistent with electroweak precision constraints for the high-mass point, while the low-mass point would require an $\mathcal{O}(1)$ tuning with the UV physics.  We have demanded in both cases that the annihilation channel dominates over all others ($>80\%$). Notice that in order to have $g_{\chi}<1$ for the annihilations into $b\bar{b}$, we must have $m_{\chi}\gtrsim 65$ GeV. The connection between the necessary freeze-out cross-section and the cross-section that explains the GCE can be made straightforwardly in the regions absent of resonance effects, due to the dominant s-wave component of the cross-sections. All points in~\figref{fig:GCE}  satisfy the correct DM relic density as measured by the Planck experiment~\cite{1502.01582}.

Though no analysis has been performed studying DM annihilation into $Zh$, and such a fit goes beyond the scope of this paper, we would like to point out that this may be an interesting annihilation channel that can most likely also explain the GCE. This is based on the fact that both annihilations into pairs of $Z$-gauge bosons and pairs of Higgses provide good fits to the GCE~\cite{Calore:2014nla} (in particular annihilation into Higgs pairs). Thus we believe that a combination of them would also provide a good fit in the approximate range of DM masses $m_{\chi}\in[108, 160]$ GeV, where annihilation into $Zh$ dominates the total cross-section. Demanding cross-sections around (1-5)$\times 10^{-26}\; \mathrm{cm}^3/\mathrm{s}$ , we find that for $0.65\lesssim g_{\chi} \lesssim 1$ and for DM masses in the range $m_{\chi}\in[108, 140]$ we expect to be able to fit the GCE and obtain at the same time the correct DM relic density.

\section{Conclusion}\label{sec:conc}

Extending the Standard Model by an additional $U(1)'$ gauge symmetry provides an economical way to address potential signals that are difficult to reconcile in the context of the SM alone. We have considered the case in which the right-handed top quark is the only SM state charged under the exotic $U(1)$. The corresponding $Z'$ gauge boson is then top-philic. The $U(1)'$ is clearly anomalous and requires the presence of spectator fermions at the scale $\Lambda_{UV}$. Rather than committing to a particular UV completion, we have taken a more general approach and considered the low energy effective description where the only new states are the $Z'$ and a Dirac fermion charged under the $U(1)'$ which is our dark matter candidate.

Previous studies considering a top-philic $Z'$ have largely focused on masses above the $t\bar{t}$ threshold. We have considered lighter $Z'$ masses in the range 150-450~GeV, which leads to a richer phenomenology. The $Z'$ can undergo kinetic mixing with the Standard Model hypercharge gauge boson. However, the mixing is constrained to be small by electroweak precision measurements. The dominant production mechanisms for the $Z'$ are then the tree-level top quark associated production ($t\bar{t}Z'$) and the loop-induced $Z'+\,$jets process.

The $Z'$ phenomenology can be divided into three distinct regions based on the dominant decays of the $Z'$.  In the low-mass region ($150\lesssim m_{Z'} \lesssim 220\,$GeV) the $Z'$ decays dominantly into $b\bar{b}$. This region can be probed by searches for $ttZ'$ associated production where the $Z'$ subsequently decays to $b\bar{b}$. Other final states, such as $Z\gamma$, can also provide constraints despite their suppressed branching ratios. Decays to $Zh$ dominate in the intermediate mass range ($220 \lesssim m_{Z'} \lesssim 300\,$GeV). This region is already strongly constrained by existing resonance searches. In the high mass region ($m_{Z'}\gtrsim 300\,$GeV) the $Z'$ decays dominantly into $t\bar{t}^{(*)}$ and the most stringent constraints arise from searches for four top quark production. We find that with $300\,\mathrm{fb}^{-1}$ at $\sqrt{s}=13\,$TeV the LHC will be able to exclude $Z'$ masses in the range 150-450~GeV for couplings $g_t\gtrsim0.2$. However, the collider bounds can be significantly weakened below the $t\bar{t}$ threshold if the $Z'$ is allowed to decay invisibly, for example to the dark matter candidate.

The combined ATLAS and CMS Higgs measurements currently show a $2.3\sigma$ excess in $t\bar{t}h$ production, driven by the same-sign dilepton channel. We find that including a contribution from $ttZ'$ associated production yields an improved fit to the data and could provide an explanation of the excess for $m_Z'\gtrsim 300\,$GeV and $g_t\gtrsim0.8$. Interestingly the lepton+jets search for four top quark production, which is sensitive to the relevant region of parameter space, has also observed a mild excess. 

The inclusion of a SM singlet Dirac fermion, charged only under the $U(1)'$, also provides a viable dark matter candidate. It is stabilised by a residual $\mathbb{Z}_2$ symmetry after the $U(1)'$ is spontaneously broken. The dark matter interacts with the SM via the $Z'$ portal and depending on the dark matter mass several channels can dominate the annihilation. We show that production via standard thermal freeze out can yield the correct relic density for a range of dark matter and $Z'$ masses. 

The Fermi-LAT collaboration has observed an excess of $\gamma$-rays emanating from the galactic center. While there are astrophysical explanations for the source of the excess photons, dark matter annihilation also provides a good fit to the observed spectrum.  We find that our dark matter candidate can provide an explanation for the excess via annihilation into $b\bar{b}$ or $t\bar{t}$ while simultaneously satisfying the relic density constraint. We anticipate that annihilation into $Zh$ may also be able to explain the excess, however currently no fit has been performed studying this final state. Finally, we find regions in parameter space which can simultaneously provide an explanation for the galactic center excess, give an improved fit to the $tth$ data, and satisfy all current collider constraints. This region of parameter space will be probed during the 13~TeV run of the LHC.

\section*{Acknowledgements}

We thank Marco Taoso, Carlos Wagner for helpful discussions. TSR acknowledges the ISIRD Grant, IIT Kharagpur, India. This work was supported by IBS under the project code, IBS-R018-D1. This work has been supported in part by the European Research Council (ERC) Advanced Grant Higgs@LHC. This work was supported by the Australian Research Council.

\appendix

\section{$Z'$ Decay Widths}\label{app:dwidth}

For reference, we list the expressions for the $Z'$ decay widths here. Loop integrals are written in terms of Passarino-Veltman functions~\cite{PVFunc}. We note that the width to $\gamma\gamma$ is identically zero by the Yang-Landau theorem~\cite{Yang:1950rg,Landau:1948kw}, while the width to $hh$ vanishes by Bose symmetry.

\subsection{Tree-Level Decays}

\subsubsection{$Z' \to t\bar{t}$, including off-shell decays}

The only SM state the $Z'$ directly couples to is the right-handed top. Below threshold, this translates to decays to $tWb$ and $WWbb$. We can express these decays through two functions, the standard triangle function $I$ defined in \modeqref{eq:Idef} and a related function $F$:
\begin{equation}
  F(a, b, c) = - \sqrt{I(a, b, c)} \, \bigl(I(a, b, c) - 3a (a - b - c)\bigr) \,.
\end{equation}
We then find that the total decay width to $WWbb$ is given by
\begin{align}
  \Gamma (Z' \to WWbb) & = \frac{\alpha_t \alpha^2}{(32\pi)^2} \, \frac{m_{Z'}}{\sin^4\theta_W} \, \frac{m_t^4}{m_W^4} \int_{q_0}^{q_1} \frac{dQ_1^2}{(Q_1^2)^2} \int_{q_0}^{q_2(Q_1^2)} \frac{dQ_2^2}{(Q_2^2)^2} \notag \\
  & \qquad \times \frac{1}{m_{Z'}^6} \, F (m_{Z'}^2 , Q_1^2, Q_2^2) \, \frac{F (m_W^2, Q_1^2, m_b^2)}{(Q_1^2 - m_t^2) + m_t^2 \Gamma_t^2} \, \frac{F (m_W^2, Q_2^2, m_b^2)}{(Q_2^2 - m_t^2) + m_t^2 \Gamma_t^2} \,.
\end{align}
The integration limits are 
\begin{equation}
  q_0 = (m_W + m_b)^2 \,, \quad q_1 = (m_{Z'} - m_W - m_b)^2 \,, \quad q_2 (x) = (m_{Z'} - \sqrt{x})^2 \,.
\end{equation}
The integration variables $Q_i^2$ are the momenta of the intermediate top quarks. When one or both top quarks are on-shell, we can use the narrow width approximation
\begin{equation}
  \frac{1}{(Q^2 - m_t^2)^2 + m_t^2 \Gamma_t^2} \approx \frac{\pi}{m_t \Gamma_t} \, \delta (Q^2 - m_t^2) \,,
\end{equation}
together with the expression for the top width
\begin{equation}
  \Gamma_t = - \frac{\alpha}{16} \frac{F (m_W^2 , m_t^2, m_b^2)}{\sin^2\theta_W \,m_t^3 m_W^2} \,.
\end{equation}

\subsubsection{$Z' \to$ leptons/light quarks}

These decays proceed through the kinetic and mass mixing only.  The decay widths for kinetic mixing  have been computed in~\cite{Chang:2006fp}. It is convenient to define two dimensionless mixings,
\begin{align}
  \kmix_Q & = \kmix \biggl( 1 - \frac{m_W^2}{m_{Z'}^2} \biggr) + \frac{\dm \sin \theta_W}{m_{Z'}^2} \,, \label{eq:defeQ}\\
  \kmix_L & = \epsilon + \frac{\dm}{m_{Z'}^2\sin \theta_W} \,. \label{eq:defeL}
\end{align}
The effective $Z'$-fermion coupling is
\begin{equation}
  \lag_{Z'f\bar{f}} = \frac{e}{\cos\theta_W} \, \frac{m_{Z'}^2}{m_{Z'}^2 - m_Z^2} \, Z'_\mu \, \bar{f} \gamma^\mu \bigl( - \kmix_Q Q^f + \kmix_L T_3^f P_L \bigr) f \,,
\label{eq:Zptoff}\end{equation}
where $\theta_W$ is the Weinberg angle, and $Q^f$ and $T_3^f$ the fermion charge and isospin respectively. Neglecting final state masses, these widths are
\begin{equation}
  \Gamma (Z' \to f\bar{f}) = \frac{N_c \alpha}{6 \cos^2\theta_W} \, m_{Z'} \, \biggl( \frac{m_{Z'}^2}{m_{Z'}^2 - m_Z^2} \biggr)^2 \, \bigl[ \kmix_Q^2 (Q^f)^2 + (\kmix_Q Q^f - \kmix_L T^f_3)^2 \bigr] \,.
\end{equation}

\subsection{Loop-Only Decays}

There are three decay modes that have no tree-level contribution even in the presence of non-zero mixing: the decays to $h\gamma$, $Z\gamma$ and $ZZ$. All the loop integrals here are finite. We define $\mu_i \equiv m_i/m_{Z'}$, $s_W \equiv \sin \theta_W$ and $c_W \equiv \cos\theta_W$. The decay widths are then
\begin{align}
  \Gamma (Z' \to h\gamma) & =  \frac{\alpha \alpha_t y_t^2}{12 \pi^3} \, \frac{m_t^2}{m_{Z'}^2 - m_h^2} \, \bigl\lvert \bigl( B_0 (m_{Z'}^2; m_t^2, m_t^2) - B_0 (m_h^2; m_t^2, m_t^2) \bigr) \notag \\
  & \qquad + 2 + (m_{Z'}^2 - m_h^2 + 4m_t^2) C_0 (0, m_{Z'}^2, m_h^2; m_t^2, m_t^2, m_t^2) \bigr\rvert^2 \,, \\
  \Gamma (Z' \to Z \gamma) & = \frac{\alpha^2 \alpha_t}{24\pi^2 s_W^2 c_W^2} \, \frac{m_{Z'}^3}{m_Z^2} \, (1 - \mu_Z^4) \, \biggl\lvert\bigl( 1 - \mu_Z^2 \bigr) m_t^2 C_0 (0, m_{Z'}^2, m_Z^2; m_t^2, m_t^2, m_t^2) \notag \\
  & \qquad + \frac{4}{3} \, s_W^2 \biggl( 1 + \frac{2 m_t^2 m_Z^2}{m_{Z'}^2} \, C_0 (0, m_{Z'}^2, m_Z^2; m_t^2, m_t^2, m_t^2) \notag \\
  & \qquad + \frac{m_Z^2}{m_{Z'}^2 - m_Z^2} \, \bigl( B_0 (m_{Z'}^2; m_t^2, m_t^2) - B_0 (m_Z^2; m_t^2, m_t^2) \bigr) \biggr)\biggr\rvert^2 \,, \\
  \Gamma (Z' \to ZZ) & = \frac{6 \alpha^2 \alpha_t}{(16\pi)^2  s_W^4 c_W^4} \, \frac{m_{Z'}^3}{m_Z^2} \sqrt{1 - 4 \mu_Z^2} \biggl\{ (1 + 2 \mu_Z^2) \frac{64}{81} \, s_W^4 + \abs{\lpint_1}^2 \biggr\} \,. 
\end{align}
In the last case we have defined the function
\begin{align}
    \lpint_1 & = \mu_t^2 \bigl\{ B_0 (m_{Z'}^2; m_t^2, m_t^2) - B_0 (m_Z^2; m_t^2, m_t^2) + m_Z^2 C_0(m_Z^2, m_{Z'}^2, m_Z^2; m_t^2, m_t^2, m_t^2) \bigr\} \notag \\
  & \qquad + \frac{4}{3} \, s_W^2 m_t^2 (1 - 4 \mu_Z^2) C_0(m_Z^2, m_{Z'}^2, m_Z^2; m_t^2, m_t^2, m_t^2) + \frac{8}{9} \, s_W^2 \notag \\
  & \qquad + \frac{16}{9} \, \mu_Z^2 s_W^2 \biggl\{ \frac{m_{Z'}^2 + 2m_Z^2}{m_{Z'}^2 - 4m_Z^2} \bigl\{ B_0 (m_{Z'}^2; m_t^2, m_t^2) - B_0 (m_Z^2; m_t^2, m_t^2) \bigr\} - 1 \notag \\
  & \qquad + \biggl( 2 m_t^2  + 2 m_Z^2 \frac{m_{Z'}^2 - m_Z^2}{m_{Z'}^2 - 4m_Z^2} \biggr) C_0(m_Z^2, m_{Z'}^2, m_Z^2; m_t^2, m_t^2, m_t^2) \biggr\} \,.
\end{align}

\subsection{Decays with Loop and Mixing Contributions}

\subsubsection{$Z' \to WW$}

The loop integrals here are finite, and the interference between the two contributions is constructive for positive kinetic mixing. We define
\begin{align}
  c_{mix} & = \tan \theta_W \frac{m_{Z'}^2}{m_{Z'}^2 - m_Z^2} \, \biggl( \kmix + \frac{\dm}{m_{Z'}^2 \sin\theta_W} \biggr) \,, \\ 
  c_{loop} & = \frac{e g_t}{(4\pi)^2s_W} \, \frac{3 m_t^2}{m_{Z'}^2 - 4m_W^2} \, \bigl[ B_0 (m_W^2; m_b^2, m_t^2) - B_0 (m_{Z'}^2; m_t^2, m_t^2) \notag \\
  & \qquad + (m_t^2 - m_b^2 - m_W^2) C_0 (m_W^2 , m_{Z'}^2 , m_W^2; m_b^2, m_t^2, m_t^2) \bigr]\,.
\end{align}
The width including interference is
\begin{multline}
  \Gamma (Z' \to WW) = \frac{\alpha m_{Z'}}{12 s_W^2} \, \frac{m_{Z'}^2}{m_W^2} \bigl(1 - \mu_W^2\bigr)^{3/2} \biggl[ \abs{c_{loop}}^2 (2 - \mu_W^2) - c_{mix} \Re (c_{loop}) \mu_W^2 (5 + 6 \mu_W^2) \\
  + \frac{1}{4} \mu_W^2 c_{mix}^2 (1 + 20\mu_W^2 + 12 \mu_W^4 ) \biggr] \,.
\end{multline}

\subsubsection{$Z' \to Zh$}

This is one of two UV-sensitive decay modes. In terms of Passarino-Veltman functions, the divergence will appear as $B_0$ functions. We define
\begin{equation}
  \bar{B}_0 (p_{10}^2, m_0^2, m_1^2) = \int_0^1 dx \log \frac{\Lambda^2}{x (1-x) p_{10}^2 + x \, m_0^2 + (1 - x) m_1^2 - i \epsilon} \,,
\end{equation}
which is $B_0$ renormalized by $\overline{MS}$ and using $\Lambda$ to represent the UV cut-off. We also define the kinematic function
\begin{equation}
  v_{hZ}^2 = I \bigl(1, \mu_h^2, \mu_Z^2 \bigr) \,.
\end{equation}
There are two separate terms in the matrix element, so we define the functions
\begin{align}
  \mathcal{A}_1 & = 3 \frac{y_t g_t}{(4\pi)^2 \sqrt{2}} \, \lpint_2 - 2 \sin\theta_W \frac{m_Z}{m_t} \, \frac{m_{Z'}^2}{m_{Z'}^2 - m_Z^2} \, \kmix_L \,, \\
  \mathcal{A}_2 & = 3 \frac{y_t g_t}{v_{hZ}^2 (4\pi)^2 \sqrt{2}} \, \lpint_3 \,,
\end{align}
with $\kmix_L$ as defined in \modeqref{eq:defeL}, and the loop integral functions
\begin{align}
  \lpint_2 & = \bar{B}_0 (m_{Z'}^2; m_t^2, m_t^2) + \bar{B}_0 (m_Z^2; m_t^2, m_t^2) - \frac{8}{3} \, s^2_W + \biggl\{ - \mu_h^2 + 4 \mu_t^2 \notag \\
  & \quad - \frac{4}{3} \, s^2_W \bigl( 1 - \mu_h^2 + \mu_Z^2 + 4 \mu_t^2 + 4 v_{hZ}^{-2} \mu_Z^2 \mu_h^2 \bigr) \biggr\} m_{Z'}^2 C_0 (m_Z^2, m_{Z'}^2, m_h^2; m_t^2, m_t^2, m_t^2) \notag \\
  & \quad - \frac{8}{3} \, s^2_W \big\{ \bigl(1 + v_{hZ}^{-2} \mu_h^2 (1 - \mu_h^2 + \mu_Z^2) \bigr) \bigl( B_0 (m_{Z'}^2; m_t^2, m_t^2) - B_0 (m_h^2; m_t^2, m_t^2) \bigr) \notag \\
  & \quad + v_{hZ}^{-2} \mu_Z^2 (1 + \mu_h^2 - \mu_Z^2) \bigl( B_0 (m_{Z'}^2; m_t^2, m_t^2) - B_0 (m_Z^2; m_t^2, m_t^2) \bigr) \bigr\} , 
\end{align}
and
\begin{align}
  v_{hZ}^2 \lpint_3 & = (1 - \mu_h^2 + \mu_Z^2) m_h^2 C_0 (m_Z^2, m_{Z'}^2, m_h^2; m_t^2, m_t^2, m_t^2) + (1 + \mu_h^2 - \mu_Z^2) \bigl( B_0 (m_{Z'}^2; m_t^2, m_t^2) \notag \\
  & \quad - B_0 (m_h^2; m_t^2, m_t^2) \bigr) + (1 - \mu_h^2 - \mu_Z^2) \bigl( B_0 (m_{h}^2; m_t^2, m_t^2) - B_0 (m_Z^2; m_t^2, m_t^2) \bigr) \notag \\
  & \quad + \frac{2}{3} \, s^2_W \bigl\{ 2 (1 - \mu_ h^2 + \mu_Z^2) \bigl[ 2 + \bigl( 1 - \mu_h^2 + \mu_Z^2 + 4 \mu_t^2 \notag \\
  & \quad\quad + 12 v_{hZ}^{-2} \mu_Z^2 \mu_h^2 \bigr) m_{Z'}^2 C_0 (m_Z^2, m_{Z'}^2, m_h^2; m_t^2, m_t^2, m_t^2)\bigr] \notag \\
  & \quad + 4 \bigl(1 + 6 v_{hZ}^{-2} \mu_Z^2 (1 + \mu_h^2 - \mu_Z^2) \bigr) \bigl( B_0 (m_{Z'}^2; m_t^2, m_t^2) - B_0 (m_h^2; m_t^2, m_t^2) \bigr) \notag \\
  & \quad + 4 \mu_Z^2 \bigl( 1 - 6 v_{hZ}^{-2} (1 - \mu_h^2 - \mu_Z^2) \bigr) \bigl( B_0 (m_{Z}^2; m_t^2, m_t^2) - B_0 (m_h^2; m_t^2, m_t^2) \bigr) \bigr\} \,.
\end{align}
The total width is
\begin{equation}
  \Gamma (Z' \to hZ) = \frac{\alpha m_{Z'}}{48 s_W^2 c^2_W} \, \frac{m_t^2}{m_Z^2} \, v_{hZ}  \Bigl[ \abs{\mathcal{A}_4}^2 \bigl(v_{hZ}^2 + 12 \mu_Z^2 \bigr) + \abs{\mathcal{A}_5}^2 v_{hZ}^2 + 2 \Re (\mathcal{A}_4 \mathcal{A}_5^{\ast}) \, v_{hZ}^2 \bigl( 1 - \mu_h^2 + \mu_Z^2 \bigr) \Bigr] \,.
\end{equation}

\subsubsection{$Z' \to b\bar{b}$}

The structure of this decay is the most complex among the leading channels. We define three functions corresponding to distinct terms in the amplitude,
\begin{align}
  \mathcal{A}_3 & = \frac{1}{3} \, \kmix_Q \tan \theta_W \, \frac{m_{Z'}^2}{m_{Z'}^2 - m_Z^2}\\
  \mathcal{A}_4 & = - \frac{g_t g_w}{32\pi^2} \, \frac{m_t^2}{m_W^2} \, \lpint_4 + T^3_b \kmix_L \tan \theta_W \, \frac{m_{Z'}^2}{m_{Z'}^2 - m_Z^2} \\
  \mathcal{A}_5 & = - \frac{g_t g_w}{32\pi^2} \, \frac{m_t^2}{m_W^2} \, \lpint_5 \,.
\end{align}
We have again used $\kmix_{L,Q}$ from Eqs.~\eqref{eq:defeQ} and~\eqref{eq:defeL}.  We have introduced two new functions, $\lpint_{4,5}$. These are loop integrals
\begin{align}
  \lpint_4 & = \frac{m_t^2 - m_b^2 -m_W^2}{m_{Z'}^2 - 4m_b^2} \bigl[ B_0 (m_b^2; m_W^2, m_t^2) - B_0 (m_{Z'}^2; m_t^2, m_t^2) \notag \\
  & \quad + (m_t^2 - m_b^2 -m_W^2) \, C_0 (m_b^2, m_{Z'}^2, m_b^2; m_W^2, m_t^2, m_t^2) \bigr] \notag \\
  & \quad - \frac{1}{2} \, \bar{B}_0 (m_{Z'}^2; m_t^2, m_t^2) + \frac{1}{2} + 2 m_W^2 \, C_0 (m_b^2, m_{Z'}^2, m_b^2; m_W^2, m_t^2, m_t^2) \,, \\
  \lpint_5 & = \frac{m_{Z'}^2}{m_{Z'}^2 - 4 m_b^2} \biggl[ 6 \frac{m_t^2 - m_b^2 -m_W^2}{m_{Z'}^2 - 4m_b^2} \bigl[ B_0 (m_b^2; m_W^2, m_t^2) - B_0 (m_{Z'}^2; m_t^2, m_t^2) \notag \\
  & \quad + (m_t^2 - m_b^2 -m_W^2) \, C_0 (m_b^2, m_{Z'}^2, m_b^2; m_W^2, m_t^2, m_t^2) \bigr] + 1 + 2 m_W^2 \, C_0 (m_b^2, m_{Z'}^2, m_b^2; m_W^2, m_t^2, m_t^2) \notag \\
  & \quad + \bigl[ B_0 (m_b^2; m_W^2, m_t^2) - B_0 (m_{Z'}^2; m_t^2, m_t^2) \bigr] - \frac{m_t^2 - m_W^2}{m_b^2} \bigl[ B_0 (m_b^2; m_W^2, m_t^2) - B_0 (0; m_W^2, m_t^2) \bigr] \biggr] \,.
\end{align}
The width is given by 
\begin{multline}
  \Gamma (Z' \to b\bar{b}) = \frac{1}{2} \, \alpha_w \, m_{Z'} \sqrt{1 - 4\mu_b^2} \Bigl[ 2 \abs{\mathcal{A}_3}^2 (1 + 2 \mu_b^2) + \abs{\mathcal{A}_4}^2 (1 - \mu_b^2) + \frac{1}{4} \abs{\mathcal{A}_5}^2 \mu_b^2 (1 - 4\mu_b^2)^2 \\
  + 2 \Re (\mathcal{A}_3 \mathcal{A}_4^\ast) (1 + 2\mu_b^2) - 2 \Re (\mathcal{A}_3 \mathcal{A}_5^\ast) \mu_b^2 (1 - 4 \mu_b^2) - \Re (\mathcal{A}_4 \mathcal{A}_5^\ast) \mu_b^2 (1 - 4\mu_b^2) \Bigr] \,.
\end{multline}

\bibliography{Zp_Phenomenology}{}
\bibliographystyle{JHEP}

\end{document}